\pdfoutput=1

\documentclass[reprint,aps,prl,twocolumn,,floatfix,superscriptaddress]{revtex4-1}
\usepackage{graphicx}
\usepackage{amssymb}
\usepackage{dcolumn}
\usepackage{bm}
\usepackage{amsmath}
\usepackage{lineno}
\usepackage[english]{babel}
\usepackage[utf8]{inputenc}
\usepackage{siunitx}
\usepackage{xcolor}
\RequirePackage[T1]{fontenc}
\RequirePackage{lmodern}

\usepackage{lineno}
\begin{document}

\title{Atomic fluctuations lifting the energy degeneracy in Si/SiGe quantum dots}

\author{Brian Paquelet Wuetz}
\thanks{These authors contributed equally}
\affiliation{QuTech and Kavli Institute of Nanoscience, Delft University of Technology, PO Box 5046, 2600 GA Delft, The Netherlands}
\author{Merritt P. Losert}
\thanks{These authors contributed equally}
\affiliation{University of Wisconsin-Madison, Madison, WI 53706 USA}
\author{Sebastian Koelling}
\thanks{These authors contributed equally}
\affiliation{Department of Engineering Physics, École Polytechnique de Montréal, Montréal, Case Postale 6079, Succursale Centre-Ville, Montréal, Québec, Canada H3C 3A7}
\author{Lucas E.A. Stehouwer}
\affiliation{QuTech and Kavli Institute of Nanoscience, Delft University of Technology, PO Box 5046, 2600 GA Delft, The Netherlands}
\author{Anne-Marije J. Zwerver}
\affiliation{QuTech and Kavli Institute of Nanoscience, Delft University of Technology, PO Box 5046, 2600 GA Delft, The Netherlands}
\author{Stephan G. J. Philips}
\affiliation{QuTech and Kavli Institute of Nanoscience, Delft University of Technology, PO Box 5046, 2600 GA Delft, The Netherlands}
\author{Mateusz T. M\k{a}dzik}
\affiliation{QuTech and Kavli Institute of Nanoscience, Delft University of Technology, PO Box 5046, 2600 GA Delft, The Netherlands}
\author{Xiao Xue}
\affiliation{QuTech and Kavli Institute of Nanoscience, Delft University of Technology, PO Box 5046, 2600 GA Delft, The Netherlands}
\author{Guoji Zheng}
\affiliation{QuTech and Kavli Institute of Nanoscience, Delft University of Technology, PO Box 5046, 2600 GA Delft, The Netherlands}
\author{Mario Lodari}
\affiliation{QuTech and Kavli Institute of Nanoscience, Delft University of Technology, PO Box 5046, 2600 GA Delft, The Netherlands}
\author{Sergey V. Amitonov}
\affiliation{QuTech and Kavli Institute of Nanoscience, Delft University of Technology, PO Box 5046, 2600 GA Delft, The Netherlands}
\author{Nodar Samkharadze}
\affiliation{QuTech and Netherlands Organisation for Applied Scientific Research (TNO), Delft, The Netherlands.}
\author{Amir Sammak}
\affiliation{QuTech and Netherlands Organisation for Applied Scientific Research (TNO), Delft, The Netherlands.}
\author{Lieven M. K. Vandersypen}
\affiliation{QuTech and Kavli Institute of Nanoscience, Delft University of Technology, PO Box 5046, 2600 GA Delft, The Netherlands}
\author{Rajib Rahman}
\affiliation{University of New South Wales, Sydney, Australia\\}
\author{Susan N. Coppersmith}
\affiliation{University of New South Wales, Sydney, Australia\\}
\author{Oussama Moutanabbir}
\affiliation{Department of Engineering Physics, École Polytechnique de Montréal, Montréal, Case Postale 6079, Succursale Centre-Ville, Montréal, Québec, Canada H3C 3A7}
\author{Mark Friesen}
\affiliation{University of Wisconsin-Madison, Madison, WI 53706 USA}
\author{Giordano Scappucci}
\email{g.scappucci@tudelft.nl}
\affiliation{QuTech and Kavli Institute of Nanoscience, Delft University of Technology, PO Box 5046, 2600 GA Delft, The Netherlands}

\date{\today}

\begin{abstract}

Electron spins in Si/SiGe quantum wells suffer from nearly degenerate conduction band valleys, which compete with the spin degree of freedom in the formation of qubits. Despite attempts to enhance the valley energy splitting deterministically, by engineering a sharp interface, valley splitting fluctuations remain a serious problem for qubit uniformity, needed to scale up to large quantum processors. Here, we elucidate and statistically predict the valley splitting by the holistic integration of 3D atomic-level properties, theory and transport. We find that the concentration fluctuations of Si and Ge atoms within the 3D landscape of Si/SiGe interfaces can explain the observed large spread of valley splitting from measurements on many quantum dot devices. Against the prevailing belief, we propose to boost these random alloy composition fluctuations by incorporating Ge atoms in the Si quantum well to statistically enhance valley splitting. 

\end{abstract}

\maketitle

Advanced semiconductor manufacturing is capable of integrating billions of transistors onto a single silicon chip. The promise of leveraging the same technology for large-scale integration of qubits into a fault-tolerant quantum processing unit is a key driver for developing electron spin qubits in silicon quantum dots\cite{vandersypen_quantum_2019}. Although these devices bear many similarities to transistors\cite{zwerver_qubits_2022}, qubits operate in the single electron regime\cite{loss_quantum_1998}, making them more sensitive to electrostatic disorder and noise arising from the surrounding environment. In strained silicon quantum wells, the electronically active part of the device is separated by an epitaxial SiGe barrier from the electronically noisy interface at the gate-stack, offering a quiet system with high mobility and low leakage between the gate and the quantum dots\cite{maune_coherent_2012}. These properties make strained Si/SiGe heterostructures promising for scalable qubit tiles\cite{Vandersypen2017InterfacingCoherent,Lieaar3960} and have made it possible to define nine quantum dot arrays\cite{zajac_scalable_2016}, run quantum algorithms\cite{Watson2018ASilicon} and entangle three-spin states\cite{takeda_quantum_2021} in natural silicon structures, and achieve two-qubit gate fidelity above 99\%\cite{xue_quantum_2022,noiri_fast_2022} in isotopically purified silicon structures. 

However, spin-qubits in silicon suffer from a two-fold degeneracy of the conduction band minima (valleys) that creates several non-computational states that act as leakage channels for quantum information\cite{zwanenburg_silicon_2013}. These leakage channels increase exponentially with the qubit count\cite{russ_theory_2020}, complicating qubit operation and inducing errors during spin transfers. Despite attempts to enhance the valley energy splitting, the resulting valley splittings are modest in Si/SiGe heterostructures, with typical values in the range of 20~to~100~\si{\micro\electronvolt}\cite{Watson2018ASilicon,zajac2015reconfigurable,shi2011tunable,scarlino2017dressed,ferdous2018valley,mi2017high,borjans2019single,mi2018landau} and only in a few instances in the range of 100~to~300~\si{\micro\electronvolt}  \cite{Borselli2011MeasurementDots,hollmann2020large,chen_detuning_2021}. Such variability in realistic silicon quantum dots remains an open challenge for scaling to large qubit systems. In particular, the probability of thermally occupying the excited valley state presents a challenge for spin initialization, and, in some cases, intervalley scattering may limit the spin coherence\cite{kawakami_electrical_2014}. Furthermore, small valley splitting may affect Pauli spin blockade readout\cite{tagliaferri_impact_2018}, which is considered in large-scale quantum computing proposals\cite{Vandersypen2017InterfacingCoherent, Lieaar3960}. Therefore, scaling up to larger systems of single-electron spin qubits requires that the valley splitting of all qubits in the system should be much larger than the typical operation temperatures ($20-100$~mK).

It has been known for some time that valley splitting depends sensitively on the interface between the quantum well and the SiGe barrier\cite{Friesen2007ValleyWells}. Past theoretical studies have considered disorder arising from the quantum well miscut angle\cite{friesen_theory_2010} and steps in the interface\cite{friesen_magnetic_2006,kharche_valley_2007, gamble2013disorder, Tariq:2019:p125309,Dodson_2022} demonstrating that disorder of this kind can greatly decrease valley splitting in quantum dots. However, a definitive connection to experiments has proven challenging for a number of reasons. At the device level, a systematic characterisation of valley splitting in Si/SiGe quantum dots has been limited because of poor device yield associated with heterostructure quality and/or device processing. At the materials level, atomic-scale disorder in buried interfaces\cite{bennett_atom_2012} may be revealed by atom-probe tomography (APT) in three-dimensions (3D) over the nanoscale dimensions comparable to electrically defined quantum dots. However, the current models employed 
to reconstruct in 3D the APT data can be fraught with large uncertainties due to the assumptions made to generate the three-dimensional representation of the tomographic data\cite{bas_general_1995}. This results in limited accuracy when mapping heterointerfaces\cite{rolland_new_2017} and quantum wells\cite{koelling_high_2009, melkonyan_atom_2017,dyck_atom_2017}. These limitations prevent linking the valley splitting in quantum dots to the relevant atomic-scale material properties and hinder the development of accurate and predictive theoretical models.

\begin{figure*}[!ht]
\includegraphics[width=183mm]{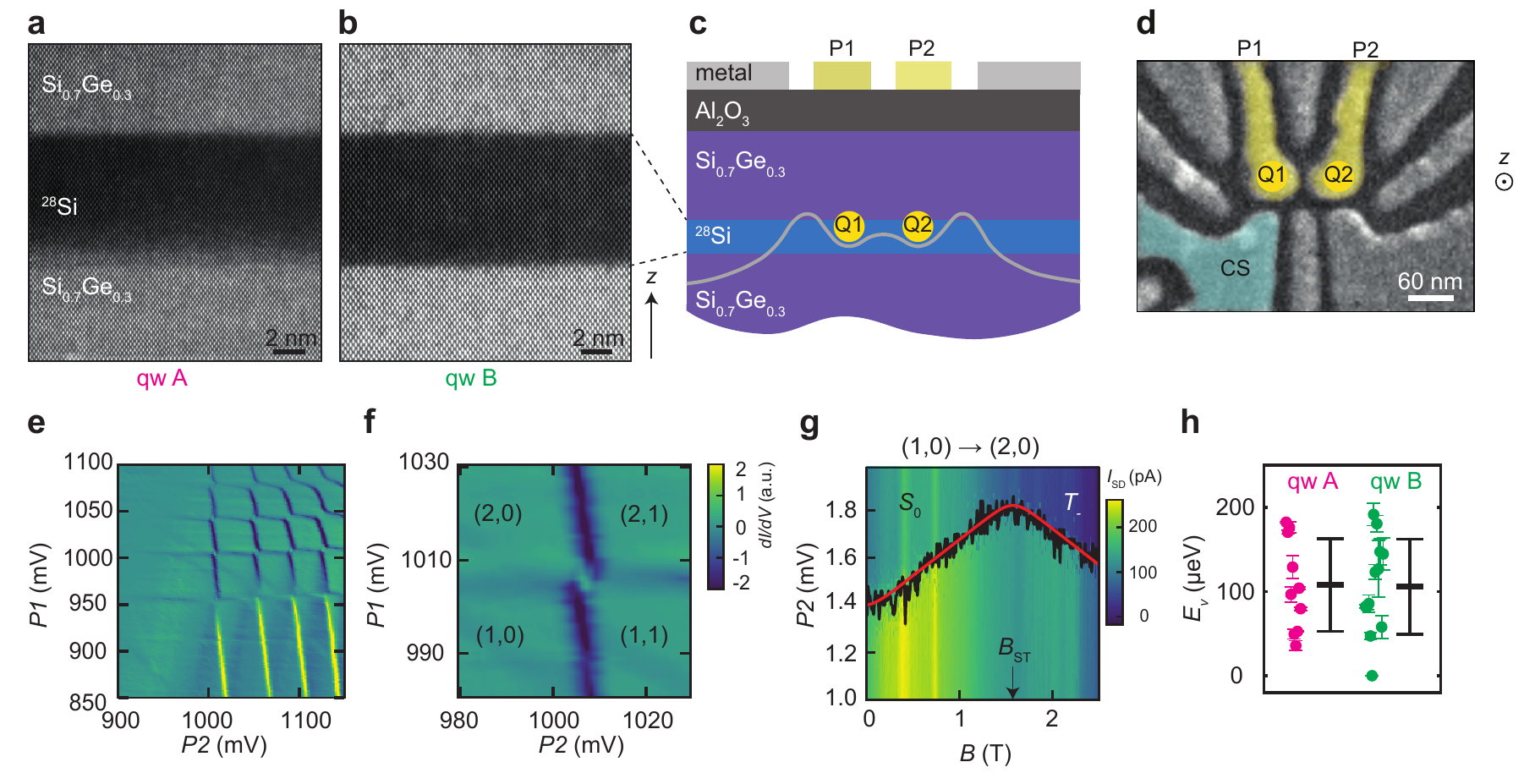}
\caption{\textbf{Material stack, devices, and valley splitting measurements} \textbf{a,b} High-angle annular dark field scanning transmission electron microscopy (HAADF-STEM) of $^{28}$Si/SiGe quantum wells A and B, respectively. \textbf{c,d} Schematic cross-section of a heterostructure with gate layout and false-coloured scanning electron microscope image of a double quantum dot, respectively. Q1 and Q2 are the quantum dots defined through confinement potentials (schematic, grey line) formed below plunger gates P1 and P2. CS is a nearby quantum dot used as a charge sensor. \textbf{e} Typical stability diagram of a double quantum dot formed by plunger gates P1 and P2 and measured by a nearby charge sensor (CS in \textbf{d}). \textbf{f} Close-up of the stability diagram in the few-electron regime. \textbf{g} Typical magnetospectroscopy of the (1,0)$\rightarrow$(2,0) transition, used to measure singlet-triplet splittings. An offset of 1082~mV is subtracted for clarity from the gate voltage applied to P2. Black lines show the location of the maximum of the differentiated charge-sensor signal ($dI_{SD}/dP2$) of the electron charging transition. Red lines show a fit to the data, from which we extract the kink position $B_\text{ST}$. The valley splitting $E_v$ is given by $g \mu_B B_\text{ST}$, where $g=2$ is the gyromagnetic ratio and $\mu_B$ is the Bohr magneton. \textbf{h} Experimental scatter plots of the valley splittings for quantum wells A (magenta) and B (green), with thick and thin horizontal black lines denoting the mean and two-sigma error bars. For quantum well B, the data point  $E_\text{V}=0$~$\mu$eV indicates that the kink in magnetospectroscopy associated with valley splitting was not observed and, consequently, that the valley splitting is below the lower bound of about 23~$\mu$eV set by our experimental measurement conditions (see Supplementary Fig.~6 and Supplementary Table 1).
}
\label{fig:1}
\end{figure*}

Herein we solve this outstanding challenge and establish comprehensive insights into the atomic-level origin of valley splitting in realistic silicon quantum dots. Firstly, we measure valley splitting systematically across many quantum dots, enabled by high-quality heterostructures with a low disorder potential landscape and by improved fabrication processes. Secondly, we establish a new method to analyse APT data leading to accurate 3D evaluation of the atomic-level properties of the Si/SiGe buried interfaces. Thirdly, incorporating the 3D atomic-level details obtained from APT, we simulate valley splitting distributions that consider the role of random fluctuations in the concentration of Si and Ge atoms at each layer of the Si/SiGe interfaces. By comparing theory with experiments, we find that the measured random distribution of Si and Ge atoms at the Si/SiGe interface is enough to account for the measured valley splitting spread in real quantum dots. Based on these atomistic insights, we conclude by proposing a practical strategy to statistically enhance valley splitting above a specified threshold as a route to making spin-qubit quantum processors more reliable --- and consequently --- more scalable.

\section{Results}
\subsection{Material stacks and devices}
Figure~\ref{fig:1} overviews the material stack, quantum dot devices, and measurements of valley splitting. To increase statistics, we consider two isotopically purified $^{28}$Si/Si$_{0.7}$Ge$_{0.3}$ heterostructures (quantum wells A and B) designed with the same quantum well width and top-interface sharpness (Methods), which are important parameters determining valley splitting\cite{Friesen2007ValleyWells,chen_detuning_2021}. As shown in high angle annular dark field scanning transmission electron microscopy (HAADF-STEM), quantum well A (Fig.~\ref{fig:1}a) has a sharp $^{28}$Si~$\rightarrow$~Si-Ge heterointerface at the top and a diffused Si-Ge~$\rightarrow$~$^{28}$Si heterointerface at the bottom, whereas in quantum well B (Fig.~\ref{fig:1}b) the growth process was optimized to achieve sharp interfaces at both ends of the quantum well. These heterostructures support a two-dimensional electron gas with high mobility and low percolation density (Supplementary Figs.~1 and 2), indicating a low disorder potential landscape, and high-performance qubits\cite{xue_cmos-based_2021,xue_quantum_2022} with single- and two-qubit gates fidelity above 99\%\cite{xue_quantum_2022}.

We define double-quantum dots electrostatically using gate layers insulated by dielectrics (Methods). A positive gate voltage applied to plunger gates P1 and P2 (Fig.~\ref{fig:1}c) accumulates electrons in the buried quantum well, while a negative bias applied to other gates tunes the confinement and the tunnel coupling between the quantum dots Q1 and Q2. All quantum dots in this work have plunger gate diameters in the range of 40-50~nm (Fig.~\ref{fig:1}d and Supplementary Table 1), setting the relevant lateral length scale for atomic-scale disorder probed by the electron wave function. 

\subsection{Valley splitting measurements}

We perform magnetospectroscopy measurements of valley splitting $E_v$ in dilution refrigerators with electron temperatures of about 100~mK (Methods). Figure~\ref{fig:1}e shows a typical charge stability diagram of a double quantum dot with DC gate voltages tuned to achieve the few electron regime, highlighted in Fig.~\ref{fig:1}f. We determine the 2-electron singlet-triplet energy splitting ($E_\text{ST}$) by measuring the gate-voltage dependence as a function of parallel magnetic field $B$ along the (0,1) $\rightarrow$ (0,2) transition (Fig.~\ref{fig:1}g) and along the (1,1) $\rightarrow$ (0,2) transition (Supplementary Fig.~4). In Fig.~\ref{fig:1}g, the transition line (black line) slopes upward, because a spin $\uparrow$ electron is added to form a singlet ground state $S_0$. Alternatively, a spin down electron can be added to form a $T_-$-state, with a downward slope.
A kink occurs when the $S_0$-state is energetically degenerate with the $T_-$-state, becoming the new ground state of the two-electron-system. From the position of the kink ($B_\text{ST}$ = 1.57~T) along the theoretical fit (red line) and the relation $E_\text{ST} = g \mu_B B_\text{ST}$, where $g=2$ is the electron gyromagnetic ratio and $\mu_B$ is the Bohr magneton, we determine $E_\text{ST}= 182.3$~\si{\micro\electronvolt} for this quantum dot. $E_\text{ST}$ sets a lower bound on the valley splitting, $E_v \geq E_\text{ST}$\cite{Borselli2011MeasurementDots,ercan2021strong}. Due to small size, our dots are strongly confined with lowest orbital energy much larger than $E_\text{ST}$ (Supplementary Fig.~3), similar to other Si/SiGe quantum dots\cite{zajac2015reconfigurable,mi2017high,hollmann2020large}. Therefore, we expect exchange corrections to have negligible effects\cite{ercan2021strong} and here take $E_{v} \approx E_\text{ST}$.

Here we report measurements of $E_{v}$ in 10 quantum dots in quantum well A and 12 quantum dots in quantum well B (Supplementary Figs.~5 and 6) and compare the measured values in Fig.~\ref{fig:1}h. We observe a rather large spread in valley splittings, however we obtain remarkably similar mean values and two-standard-deviation error bars $\overline{E_v}\pm2\sigma$ of $108\pm55$~\si{\micro\electronvolt} and $106\pm58$~\si{\micro\electronvolt} for quantum wells A and B, respectively\footnote{The quantum dots all have a similar design and hence are expected to have similar electric fields across the devices with a small influence on valley splitting under our experimental conditions}. We argue that quantum wells A and B have similar $\overline{E_v}\pm2\sigma$ because the electronic ground state is confined against the top~interface, which is very similar in the two quantum wells.

\begin{figure*}[!ht]
\includegraphics[width=183mm]{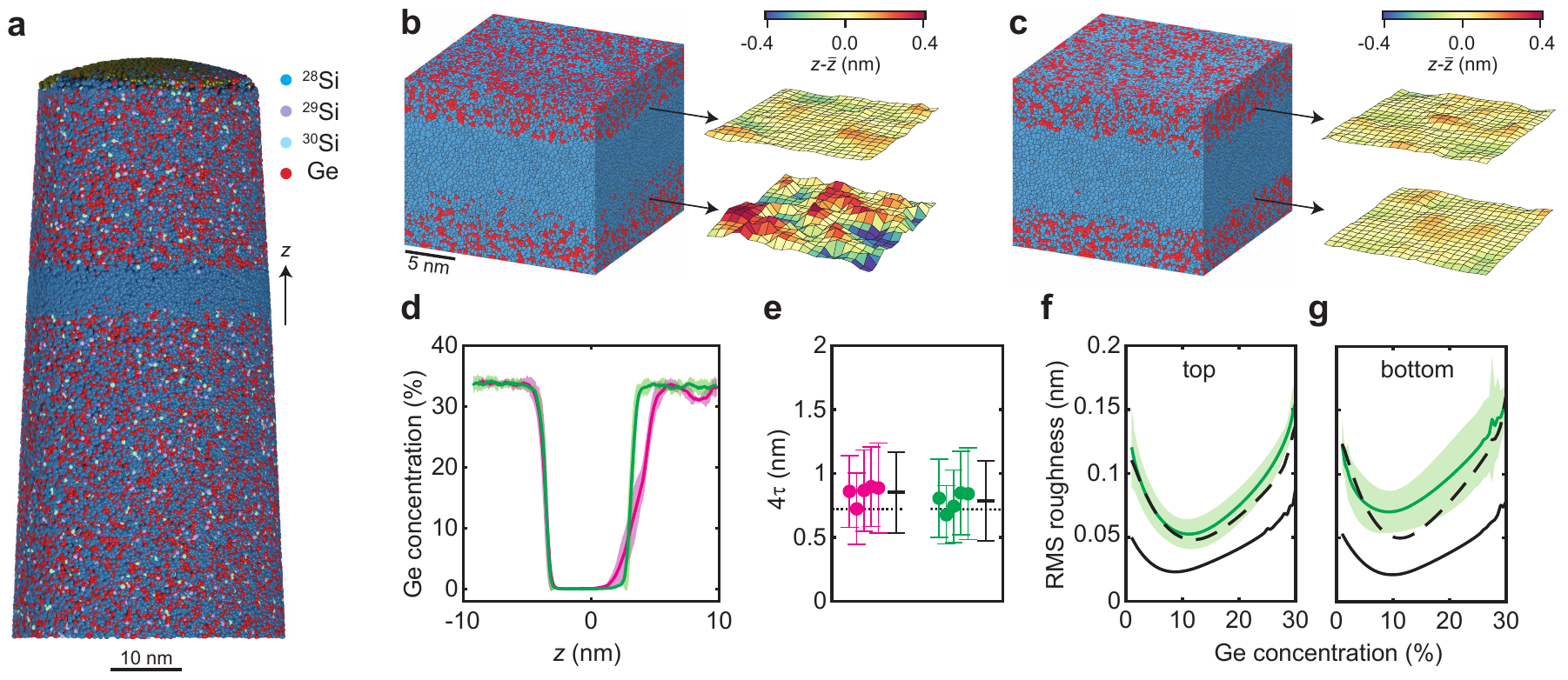}%
\caption{\textbf{Atom probe tomography of $^{28}$Si/SiGe heterostructures.} \textbf{a} Point-cloud APT reconstruction of quantum well B, showing the $^{28}$Si quantum well and surrounding SiGe barriers. Isotopic purification is confirmed by secondary ion mass spectroscopy (Supplementary Fig.~14). \textbf{b}, \textbf{c} Voronoi tessellation of the APT reconstructions for quantum wells A and B, respectively, and extracted isosurfaces corresponding to 8\% Ge concentration. $\bar z$ is the average position of the 8\% Ge concentration across these particular samples. We limit the lateral size of the analysis to $\approx 30\text{~nm}\times30\text{~nm}$, reflecting the typical lateral size of a quantum dot (Fig.~\ref{fig:1}d). \textbf{d} Average germanium concentration depth profiles across quantum wells A (magenta) and B (green). Shaded areas mark the 95\% confidence interval over each of the sets of five APT samples. \textbf{e} Statistical analysis of the top interface width $4\tau$ determined by fitting the data for quantum wells A (magenta) and B (green) to sigmoid functions. Thick and thin horizontal black lines denote the mean and two-standard-deviation error bars for the different APT samples. Dotted black lines show $4\tau$ results from the HAADF-STEM measurements (Supplementary~Fig 13). \textbf{f},\textbf{g} Root mean square (RMS) roughness of the concentration isosurfaces as a function of germanium concentration at the top and bottom interfaces of quantum well B (green line). Shaded areas indicate the 95\% confidence interval, averaged over each set of five APT samples. The experimental data are compared to the RMS roughness of a simulated quantum well with the interface properties of \textbf{d} (dashed black line) vs. an atomically sharp quantum well (solid black line).}
\label{fig2}
\end{figure*}

\subsection{Atom probe tomography}

We now characterise the atomic-scale concentration fluctuations at the quantum well interfaces to explain the wide range of measured valley splittings with informed theoretical and statistical models. To probe the concentrations over the dimensions relevant for quantum dots across the wafer, we perform APT on five samples each from quantum wells A and B, with a field of view of approximately 50~nm at the location of the quantum well (Methods). First, we show how to reliably reconstruct the buried quantum well interfaces, then we use this methodology to characterise their broadening and roughness. 

Figure~\ref{fig2}a shows a typical point-cloud reconstruction of an APT specimen from quantum well B. Each point represents the estimated position of an ionized atom detected during the experiment\cite{bas_general_1995}. Qualitatively, we observe an isotopically enriched $^{28}$Si quantum well, essentially free of $^{29}$Si, cladded in a SiGe alloy. To probe the interface properties with the highest possible resolution allowed by APT and differently from previous APT studies on Si/SiGe\cite{dyck_atom_2017}, we represent the atom positions in the acquired data sets in form of a Voronoi tessellation\cite{Voronoi+1908+97+102,Voronoi+1908+198+287} and generate profiles on an $x-y$ grid of the tessellated data, as described in Supplementary Discussion Section 2c. A sigmoid function $[1+\exp(z-z_0)/\tau]^{-1}$\cite{dyck_atom_2017} is used to fit the profiles of each tile in the $x-y$ grid. Here, $z_0$ is the inflection point of the interface and $4\tau$ is the interface width. As the Voronoi tessellation of the data set does not sacrifice any spatial information, the tiling in the $x-y$ plane represents the smallest lateral length scale over which we characterise the measured disorder at the interface. Note that we do not average at all over the $z$ axis and hence maintain the inherent depth resolution of APT. We find that for tiles as small as $3\text{~nm}\times3\text{~nm}$ the numerical fitting of sigmoid functions to the profiles converges reliably. Although each tile contains many atoms, their size is still much smaller than the quantum dot diameter, and may therefore be considered to be microscopic. We use the sigmoid fits for each tile stack to visualise and further characterise the interfaces (Supplementary Figs.~8--10). Importantly, Ge concentration isosurfaces as shown in Fig.~\ref{fig2}b,c are constructed by determining the vertical position for which each of the sigmoids reaches a specific concentration. Note, that we oversample the interface to improve the lateral resolution by making the $3\text{~nm}\times3\text{~nm}$ tiles partially overlap (Supplementary Discussion Section 2c).

In Fig.~\ref{fig2}d, we show the average Ge concentration profile and measurement to measurement variations from the tessellated volumes (Supplementary Discussion Section 2b,c) of all samples for both quantum wells A and B. APT confirms HAADF-STEM results in Fig.~\ref{fig:1}a,b: quantum wells A and B have an identical sharp top interface and quantum well A has a broader bottom interface. Furthermore, the shaded colored areas in Fig.~\ref{fig2}d reveal narrow 95\% confidence levels, pointing to highly uniform concentration profiles when averaged across the wafer. Strong disorder fluctuations emerge at the much smaller tile length scale. In Fig.~\ref{fig2}e we show for all samples of a given quantum well the interface width mean value with two standard deviations $\overline{4\tau }\pm2\sigma$, obtained by averaging over all the tiles in a given sample. The results indicate uniformity of $\overline{4\tau}$, and further averaging across all samples of a given heterostructure ($\mu_{\overline{4\tau}}$, black crosses) yields similar values of
$\mu_{\overline{4\tau}}
=0.85\pm0.32$~nm and $0.79\pm0.31$~nm for quantum wells A and B, consistent with our $4\tau$ analysis from HAADF-STEM measurements (black dotted lines). However, the two-standard-deviation errors ($2\sigma$) of each data point can be up to 30\% of the mean value $\overline{4\tau}$. 

To pinpoint the root cause of atomic-scale fluctuations at the interface, in Fig.~\ref{fig2}f,g we utilize the 3D nature of the APT data sets, calculate, and compare the root mean square (RMS) roughness of the interfaces (solid green lines) as measured by APT on quantum well B to two 3D models (Fig.~\ref{fig2}f,g) mimicking the dimensions of an APT data set. Both models are generated with random distributions of Si and Ge in each atomic plane (Supplementary Discussion Section~2d). The first model (solid black lines) corresponds to an atomically abrupt interface where the Ge concentration drops from $\sim$33.5\% to 0\% in a single atomic layer. It hence represents the minimum roughness achievable at each isoconcentration surface given the in-plane randomness of SiGe and the method to construct the interface. The second model (dashed black lines) is generated with the experimentally determined Ge concentration profile along the depth axis (Supplementary Fig.~11). As shown in Fig.~\ref{fig2}f,g, the roughness extracted from the second model fits well to the measured data, suggesting that the RMS roughness measured by APT is fully explained by the interface width and shape along the depth axis. Furthermore, as the deviation of each isosurface tile position from the isosurface's average position also matches that of the measured interfaces from the second model (Supplementary~Movie~1) the APT data are consistent with a random in-plane distribution of Ge perpendicular to the interface in all data sets of quantum well B. For 2 out of 5 samples on quantum well A that we analyzed, we observe features that are compatible with correlated disorder from atomic steps (Supplementary Fig.~13). In the following, the alloy disorder observed in the APT concentration interfaces is incorporated into a theoretical model. As shown below, the calculations of valley splitting distributions associated with the 3D landscape of Si/SiGe interfaces can be further simplified into a 1D model that incorporates the in-plane random distribution of Si and Ge atoms. 

\begin{figure*}[!ht]
	\includegraphics[width=183mm]{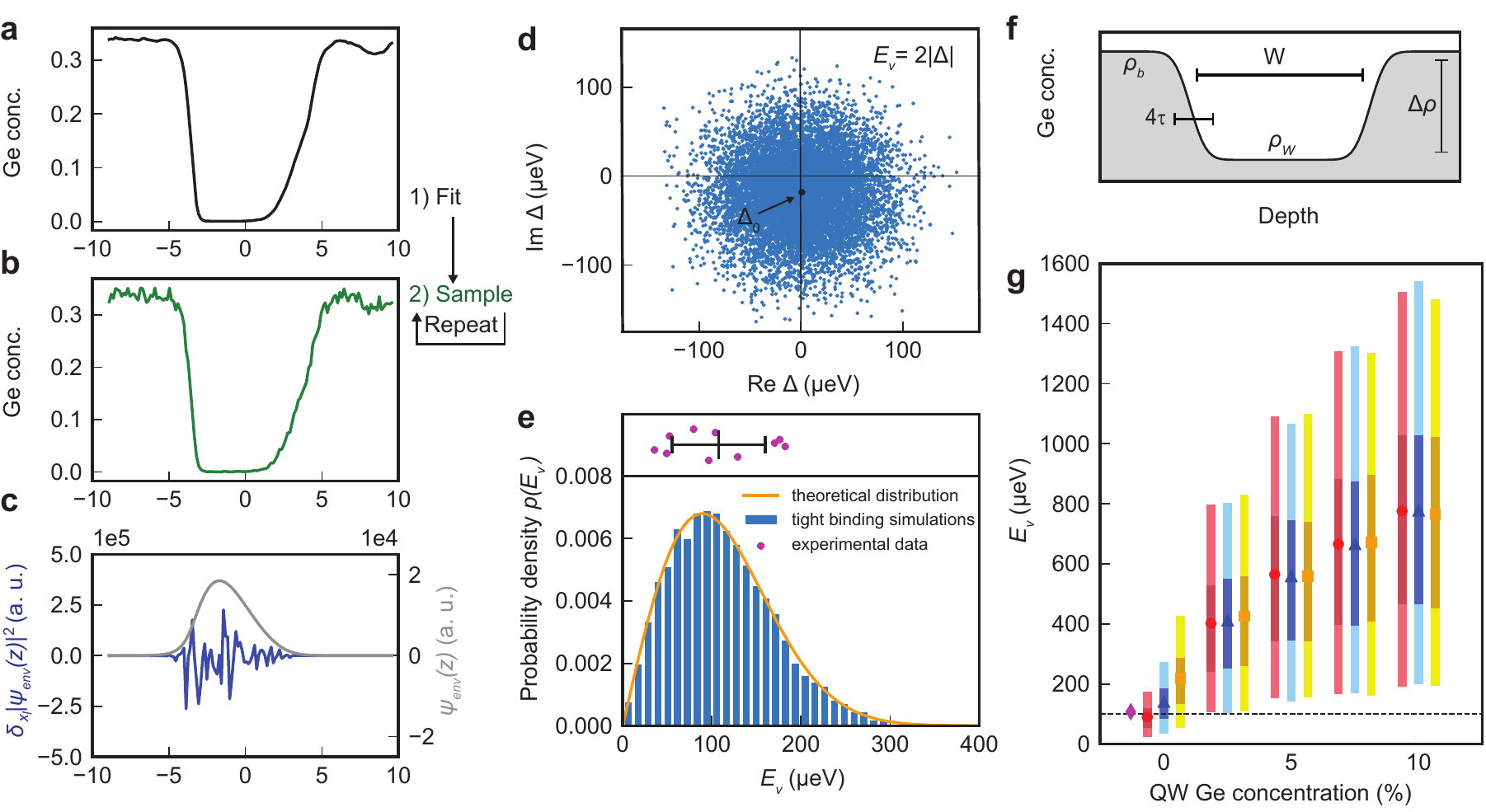}%
	\caption{\textbf{Valley-splitting simulations.} \textbf{a} Average concentration profile obtained from APT data (quantum well A). \textbf{b}, Typical, randomized Ge concentration profile, derived from \textbf{a}. \textbf{c} Envelope function $\psi_\text{env}(z)$, obtained for the randomized profile in \textbf{b} (grey curve), and the corresponding concentration fluctuations weighted by the envelope function squared: $\delta_{x_l} |\psi_\text{env}(z_l)|^2$ (blue). Here, the wavefunction is concentrated near the top interface where the concentration fluctuations are also large; the weighted fluctuations are therefore largest in this regime. \textbf{d} Distribution of the intervalley matrix element $\Delta$ in the complex plane, as computed using an effective-mass approach, for 10,000 randomized concentration profiles. The black marker indicates the deterministic value of the matrix element $\Delta_0$, obtained for the experimental profile in \textbf{a}. \textbf{e}, Histogram of the valley splittings from tight-binding simulations with 10,000 randomized profiles. The same profiles may be used to compute valley splittings using effective-mass methods; the orange curve shows a Rice distribution whose parameters are obtained from such effective-mass calculations (see Methods).
	\textbf{f}, Schematic Si/SiGe quantum well with Ge concentrations $\rho_W$ (in the well) and $\rho_b = \rho_W+\Delta \rho$ (in the barriers), with a fixed concentration difference of $\Delta \rho = 25$\%. \textbf{g}, Distribution of valley splittings obtained from simulations with variable Ge concentrations, corresponding to $\rho_W$ ranging from 0 to 10\%, and interface widths $4\tau=5$~ML (red circles), 10~ML (blue triangles), or 20~ML (orange squares), where ML refers to atomic monolayers. Here, the marker describes the mean valley splitting, while the darker bars represent the 25-75 percentile range and the lighter bars represent the 5-95 percentile range. Each bar reflects 2,000 randomized tight-binding simulations of a quantum well of width $W = 120$~ML. The magenta diamond at zero Ge concentration shows the average measured valley splitting of quantum well A. In all simulations reported here, we assume an electric field of $E = 0.0075$~V/nm and a parabolic single-electron quantum-dot confinement potential with orbital excitation energy $\hbar \omega = 4.18$~meV and corresponding dot radius $\sqrt{\hbar/m^*\omega}$.
	}
\label{fig:theory}
\end{figure*}

\subsection{Valley splitting simulations}

We begin by considering an ideal laterally infinite heterostructure with no concentration fluctuations, and we denote the average Si concentration at layer $l$ by $\bar x_l$. Due to the finite size of a quantum dot and the randomness in atomic deposition, there will be dot-to-dot concentration fluctuations. We therefore model the actual Si concentration at layer $l$ by averaging the random alloy distribution weighted by the lateral charge density in the quantum dot, giving $x^d_l = \bar x_l + \delta_{x_l}$, as described in Supplementary Discussion Section 3c. Here, the random variation $\delta_{x_l}$ is computed assuming a binomial distribution of Si and Ge atoms. We find that these fluctuations can have a significant impact on the valley splitting. 

We explore these effects numerically using 1D tight-binding simulations. We begin with the averaged fitted concentration profiles obtained from the APT analysis in Fig~\ref{fig2}d, which enable us to directly measure the average Ge concentration in a given layer $\bar x_l$ (Fig.~\ref{fig:theory}a). The variance of the concentration fluctuations is determined by the size of the quantum dot, which we assume has an orbital excitation energy of $\hbar \omega = 4.18 \: \mathrm{meV}$ and corresponding radius $\sqrt{\hbar/m^*\omega}$, as well as the average Si concentration $\bar x_l$. Here, $m^*$ is the effective mass of Si. Together, $\bar x_l$ and the variance determine the probability distribution of weighted Si and Ge concentrations. Concentration profiles are sampled repeatedly from this distribution, with a typical example shown in Fig.~\ref{fig:theory}b. The valley splitting is then determined from a 1D tight-binding model~\cite{Boykin:2004:p165325}. The envelope of the effective mass wavefunction $\psi_\text{env}(z)$ is shown in Fig.~\ref{fig:theory}c (grey curve) for an electron confined in the quantum well of Fig.~\ref{fig:theory}b. The procedure is repeated for 10,000 profile samples, obtaining the histogram of valley splittings shown in Fig.~\ref{fig:theory}e. These results agree very well with calculations obtained using a more sophisticated three-dimensional 20-band sp\textsuperscript{3}d\textsuperscript{5}s* NEMO tight-binding model\cite{klimeck_atomistic_2007} (Supplementary Discussion Section 3b) and confirm that concentration fluctuations can produce a wide range of valley splittings. For comparison, at the top of Fig.~\ref{fig:theory}e, we also plot the same experimental valley splittings shown in Fig.~\ref{fig:1}h, demonstrating good agreement in both the average value and the statistical spread.
These observations support our claim that the valley splitting is strongly affected by composition fluctuations due to random distributions of Si and Ge atoms near the quantum well interfaces, even though the experiments cannot exclude the presence of correlated disorder from atomic steps in quantum dots. 

Analytical methods using effective mass theory may also be used to characterise the distribution of valley splittings. First, we model the intervalley coupling matrix element~\cite{Friesen2007ValleyWells} as $\Delta = \int e^{-2ik_0 z_l} U(z) |\psi_\text{env}(z)|^2 dz $, where $k_0 = 0.82 \cdot 2\pi/a_0$ is the position of the valley minimum in the Si Brillouin zone, $a_0 = 0.543$~nm is length of the Si cubic unit cell, $\psi_\text{env}(z)$ is a 1D envelope function, and $U(z)$ is the quantum well confinement potential. The intervalley coupling $\Delta$ describes how sharp features in the confinement potential couple the two valley states, which would otherwise be degenerate. In general, $\Delta$ is a complex number that can be viewed as the sum of two distinct components: a deterministic piece $\Delta_0$, arising from the average interface concentration profile, and a random piece $\delta \Delta$, arising from concentration fluctuations. The latter can be expressed as a sum of contributions from individual atomic layers: $\delta \Delta = \sum_l \delta \Delta_l$, where $\delta\Delta_l$ is proportional to $\delta_{x_l} |\psi_\text{env}(z_l)|^2$ (see Methods). To visualize the effects of concentration fluctuations in Fig.~\ref{fig:theory}c, we compute $\delta\Delta_l$ using the randomized density profile of Fig.~\ref{fig:theory}b (blue curve). We see that most significant fluctuations occur near the top interface, where $|\psi_\text{env} (z_l)|$ and the Ge content of the quantum well are both large. In Fig.~\ref{fig:theory}d we plot $\Delta$ values obtained for 10,000 quantum-well realizations using this effective mass approach. The deterministic contribution to the valley splitting $\Delta_0$ (black dot) is seen to be located near the center of the distribution in the complex plane, as expected. However, the vast majority of $\Delta$ values are much larger than $\Delta_0$, demonstrating that concentration fluctuations typically provide the dominant contribution to intervalley coupling.

The total valley splitting is closely related to the intervalley coupling via $E_v = 2|\Delta|$, and therefore exhibits the same statistical behavior. In Fig.~\ref{fig:theory}e, the orange curve shows the Rice distribution whose parameters are derived from effective-mass calculations of the valley splitting (see Methods), using the same concentration profiles as the histogram data. The excellent agreement between these different approaches confirms the accuracy of our theoretical techniques (Supplementary Discussion Section 3d). 

\section{Discussion}
Based on the results obtained above, we now propose two related methods for achieving large valley splittings (on average), with high yields. Both methods are derived from the key insight of Fig.~\ref{fig:theory}c: due to random-alloy fluctuations, the valley splitting is almost always enhanced when the electronic wavefunction overlaps with more Ge atoms. In the first method, we therefore propose to increase the width of the interface ($4\tau$) as shown in Fig.~\ref{fig:theory}f, since this enhances the wavefunction overlap with Ge atoms at the top of the quantum well. This approach is nonintuitive because it conflicts with the conventional deterministic approach of engineering sharp interfaces. The second method, also shown in Fig.~\ref{fig:theory}f, involves intentionally introducing a low concentration of Ge inside the quantum well. The latter method is likely more robust because it can incorporate both deterministic enhancement of the valley splitting from a sharp interface, and fluctuation-enhanced valley splitting.

We test these predictions using simulations, as reported in Fig.~\ref{fig:theory}g, where different colors represent different interface widths and the horizontal axis describes the addition of Ge to the quantum well. For no intentional Ge in the quantum well, as consistent with the heterostructure growth profile of our measured quantum dots, the calculations show a significant increases in the valley splitting with increasing interface width. Here, the narrowest interface appears most consistent with our experimental results (magenta marker), attesting to the sharp interfaces achieved in our devices. As the Ge concentration increases in the quantum well, this advantage is largely overwhelmed by concentration fluctuations throughout the well. A very substantial increase in valley splitting is observed for all concentration enhancements, even at the low, 5\% level. Here, the light error bars represent 5-95 percentiles while dark bars represent 25-75 percentiles. At the 5\% concentration level, our simulations indicate that $>$95\% of devices should achieve valley splittings $>$100~\si{\micro\electronvolt}. This value is more than an order of magnitude larger than the typical operation temperature of spin-qubits and is predicted to yield a 99\% readout fidelity\cite{tagliaferri_impact_2018}. This would represent a significant improvement in qubit yield for Si quantum dots. A recent report of SiGe quantum wells with oscillating Ge concentrations provides the first experimental evidence that intentionally placing Ge in the quantum well leads to significant variability and some of the highest recorded values of valley splitting\cite{mcjunkin_sige_2021}.

In conclusion, we argue for the atomic-level origin of valley splitting distributions in realistic Si/SiGe quantum dots, providing key insights on the inherent variability of Si/SiGe qubits and thereby solving a longstanding problem facing their scaling. We relate 3D atom-by-atom measurements of the heterointerfaces to the statistical electrical characterisation of devices, and ultimately to underlying theoretical models. We observe qualitative and quantitative agreement between simulated valley splitting distributions and measurements from several quantum dots, supporting our theoretical framework. Crucially, we learn that atomic concentration fluctuations of the $^{28}$Si~$\rightarrow$~Si-Ge heterointerface are enough to account for the valley splitting spread and that these fluctuations are largest when the envelope of the wavefunction overlaps with more Ge atoms. Moreover, while we have only incorporated random alloy disorder into our theoretical framework so far, we foresee that APT datasets including correlated disorder, such as steps, will be used to further refine our theoretical understanding of valley splitting statistics. Since atomic concentration fluctuations are always present in Si/SiGe devices due to the intrinsic random nature of the SiGe alloy, we propose to boost these fluctuations to achieve on average large valley splittings in realistic silicon quantum dots, as required for scaling the size of quantum processors. Our proposed approaches are counter-intuitive yet very pragmatic. The interface broadening approach seems viable for hybrid qubits, which require valley splitting to be large enough to be usable but not so large as to be inaccessible. For single-electron spin qubits, which don’t use the valley degree of freedom, the direct introduction of Ge in the quantum well appears better suited for targeting the largest possible valley splitting. By adding Ge to the Si quantum well in small concentrations we expect to achieve on average valley splitting in excess of 100~\si{\micro\electronvolt}. Early calculations from scattering theories\cite{monroe_comparison_1993} suggest that the added scattering from random alloy disorder will not be the limiting factor for mobility in current $^{28}$Si/SiGe heterostructures. However, an approximate twofold reduction in electron mobility was recently reported when an oscillating Ge concentration of about 5\% on average is incorporated in the Si quantum well\cite{mcjunkin_sige_2021}. We speculate that fine tuning of the Ge concentration in the quantum well will be required for enhancing the average valley splitting whilst not compromising the low-disorder potential environment, which is important for scaling to large qubit systems. We believe that our results will inspire a new generation of Si/SiGe material stacks that rely on atomic-scale randomness of the SiGe as a new dimension for the heterostructure design. 

\bibliography{bibliography.bib}

\begin{footnotesize}
\section{Methods}

\noindent\textbf{Si/SiGe heterostructure growth.} The $^{28}$Si/SiGe heterostructures are grown on a 100-mm n-type Si(001) substrate using an Epsilon 2000 (ASMI) reduced pressure chemical vapor deposition reactor equipped with a $^{28}$SiH$_4$ gas cylinder (1\% dilution in H$_2$) for the growth of isotopically enriched $^{28}$Si. The $^{28}$SiH$_4$ gas was obtained by reducing $^{28}$SiF$_4$ with a residual $^{29}$Si concentration of 0.08\%\cite{SabbaghPhysRevApplied2019}. Starting from the Si substrate, the layer sequence for quantum well A comprises a 900~nm layer of Si$_{1-x}$Ge$_x$ graded linearly from $x=0$ to 0.3, followed by a 300~nm Si$_{0.7}$Ge$_{0.3}$ strain-relaxed buffer, an 8~nm tensily strained $^{28}$Si quantum well, a 30~nm Si$_{0.7}$Ge$_{0.3}$ barrier, and a sacrificial Si cap. The layer sequence for quantum well B comprises a 1.4~$\si{\micro\meter}$ step-graded Si$_{(1-x)}$Ge$_x$ layer with a final Ge concentration of $x = 0.3$ achieved in four grading steps ($x = 0.07$, 0.14, 0.21, and 0.3), followed by a 0.45~$\si{\micro\meter}$ Si$_{0.7}$Ge$_{0.3}$ strain-relaxed buffer, an 8~nm tensily strained $^{28}$Si quantum well, a 30~nm Si$_{0.7}$Ge$_{0.3}$ barrier, and a sacrificial Si cap. 
In quantum well A, the Si$_{0.7}$Ge$_{0.3}$ strain-relaxed buffer and the Si quantum well are grown at 750~$^{\circ}$C without growth interruption. In quantum well B the Si$_{0.7}$Ge$_{0.3}$ strain-relaxed buffer below the quantum well is grown at a temperature of 625~$^{\circ}$C, followed by growth interruption and quantum well growth at 750~$^{\circ}$C. This modified temperature profile yields a sharper bottom interface for quantum well B as compared to quantum well A.

\noindent\textbf{Atom probe tomography.} Samples for APT were prepared in a FEI Helios Nanolab 660 dual-beam scanning electron microscope using a gallium focused ion beam at 30, 16 and 5~kV and using a procedure described in detail in ref.~\cite{koelling_invited_2020}. Before preparation, a 150-200~nm thick chromium capping layer was deposited on the sample via thermal evaporation to minimize the implantation of gallium ions into the sample. All APT analyses were started inside this chromium cap with the stack fully intact underneath. APT was carried out using a LEAP 5000XS tool from Cameca. The system is equipped with a laser to generate picosecond pulses at a wavelength of 355~nm. For the analysis, all samples were cooled to a temperature of 25~K. The experimental data are collected at a laser pulse rate of 200-500~kHz at a laser power of 8-10~pJ. APT data are reconstructed using IVAS~3.8.5a34 software and visualized using the AtomBlend addon to Blender~2.79b and Blender~2.92 software. For the Voronoi tessellation the reconstructed data sets were exported to Python~3.9.2 and then tessellated using the scipy.spatial.Voronoi class of SciPy~1.6.2. Note that in these analyses the interfaces are represented as an array of sigmoid functions generated perpendicular to the respective interface on $3\text{~nm}\times3\text{~nm}$ tiles that are 1~nm apart. This sacrifices lateral resolution to allow for statistical sampling of the elemental concentrations but preserves the atomic resolution along the depth axis that APT is known to provide upon constructing the interface as shown in Fig.~\ref{fig2}a.

\textbf{Device fabrication.} The fabrication process for Hall-bar shaped heterostructure field effect transistors (H-FETs) involves: reactive ion etching of mesa-trench to isolate the two-dimensional electron gas (2DEG); P-ion implantation and activation by rapid thermal annealing at 700~$^{\circ}$C; atomic layer deposition of a 10-nm-thick Al$_2$O$_3$ gate oxide; deposition of thick dielectric pads to protect gate oxide during subsequent wire bonding step; sputtering of Al gate; electron beam evaporation of Ti:Pt to create ohmic contacts to the 2DEG via doped areas. All patterning is done by optical lithography. Quantum dot devices are fabricated on wafer coupons from the same H-FET fabrication run and share the process steps listed above. Double-quantum dot devices feature a single layer gate metallization and further require electron beam lithography, evaporation of Al (27~nm) or Ti:Pd (3:27~nm) thin film metal gate, and lift-off. For linear quantum dot arrays the gate stack consists of 3 layers of Ti:Pd metallic gates (3:17, 3:27, 3:27~nm) isolated from each other by 5~nm Al$_2$O$_3$ dielectric interlayers. The fabrication processes for quantum dot devices are further detailed in ref.~\cite{lawrie2020quantum}.

\textbf{Electrical characterisation of devices.} Hall-bar measurement are performed in a Leiden cryogenic dilution refrigerator with a mixing chamber base temperature $T_\text{MC} = 50$mK\cite{paquelet_wuetz_multiplexed_2020}. We apply a source-drain bias of 100 $\si{\micro\volt}$ and measure the source-drain current $I_\text{SD}$, the longitudinal voltage $V_{xx}$, and the transverse Hall voltage $V_{xy}$ as function of the top gate voltage $V_g$ and the external perpendicular magnetic field $B$. From here we calculate the longitudinal resistivity $\rho_{xx}$ and transverse Hall resistivity $\rho_{xy}$. The Hall electron density $n$ is obtained from the linear relationship $\rho_{xy}=B/en$ at low magnetic fields. The carrier mobility $\mu$ is extracted from the relationship $\sigma_{xx}=ne\mu$, where $e$ is the electron charge. The percolation density $n_p$ is extracted by fitting the longitudinal conductivity $\sigma_{xx}$ to the relation $\sigma_{xx} \propto (n - n_p)^{1.31}$. Here $\sigma_{xx}$ is obtained via tensor inversion of $\rho_{xx}$ at $B = 0$. Quantum dot measurements are performed in Oxford and Leiden cryogenic refrigerators with base temperatures ranging from 10--50~mK. Quantum dot devices are operated in the few-electron regime. Further details of the 2DEG and quantum dot measurements are provided in the Supplementary Discussion Section~1.

\textbf{Theory and simulations.} 
The quantum-well potential at vertical position $z_l$ is simply defined here as a linear interpolation of the conduction-band offset at the quantum-well interface: $U(z_l) = \frac{x^d_l - x_s}{x_w - x_s} \Delta E_c$, where $x_l^d$ is the average Si concentration in layer $l$, $x_s$ is the average Si concentration in the strain-relaxed SiGe barriers, $x_w$ is the average Si concentration in the strained quantum well, and $\Delta E_c$ is the conduction band offset in the absence of fluctuations. In the effective-mass theory, the intervalley coupling matrix element can then be approximated by the sum
\begin{equation} \label{eq:matrixElement}
    \Delta = \frac{a_0}{4} \sum_l e^{-2ik_0 z_l} \frac{x^d_l - x_s}{x_w - x_s} \Delta E_c |\psi_\text{env}(z_l)|^2.
\end{equation}
Defining the local concentration fluctuations as $x_l^d = \bar x_l + \delta_{x_l}$, the matrix element can then be split into its deterministic and fluctuating contributions $\Delta = \Delta_0 + \delta \Delta$, where the fluctuating term $\delta \Delta$ contains all dependence on $\delta_{x_l}$:
\begin{equation} \label{eq:delta2}
   \delta \Delta = \frac{a_0}{4} \frac{\Delta E_c}{x_w - x_s} \sum_l e^{-2ik_0 z_l} \delta_{x_l} |\psi_\text{env}(z_l)|^2. 
\end{equation}
The deterministic term $\Delta_0$ represents the matrix element of the ideal, smooth concentration profile, while $\delta \Delta$ describes the fluctuations about this value. For concentration fluctuations $\delta_{x_l}$ defined by binomial distributions of Ge and Si atoms, the resulting valley splitting $E_v = 2|\Delta_0 + \delta \Delta|$ corresponds to a Rice distribution with parameters $\nu = 2|\Delta_0|$ and $\sigma = \sqrt{2} \sqrt{\mathrm{Var}\left[\delta \Delta \right]}$ \cite{Aja-Fernandez:2016}. For additional details, see the Supplementary Discussion Section 3. All simulations and numerical calculations reported in this work were performed using Python 3.7.10 with the open-source libraries NumPy, SciPy, and Matplotlib. 
The 3D atomistic simulations were done using the large-scale Slater-Koster tight-binding solver NEMO3D. A spin resolved 20 band sp3d5s* nearest neighbour model was used. Strain optimization was done using a valence force field Keating model. 

\section{Data availability}
\noindent All data included in this work are available from the 4TU.ResearchData international data repository at https://doi.org/10.4121/16592522.

\section{Acknowledgements}
\noindent This work was supported in part by the Army Research Office (Grant No. W911NF-17-1-0274). The views and conclusions contained in this document are those of the authors and should not be interpreted as representing the official policies, either expressed or implied, of the Army Research Office (ARO), or the U.S. Government. The U.S. Government is authorized to reproduce and distribute reprints for Government purposes notwithstanding any copyright notation herein.
\noindent The APT work was supported by NSERC Canada (Discovery, SPG, and CRD Grants), Canada Research Chairs, Canada Foundation for Innovation, Mitacs, PRIMA Québec, and Defence Canada (Innovation for Defence Excellence and Security, IDEaS).

\section{Author contributions}
\noindent A.S. grew and designed the Si/SiGe heterostructures with B.P.W. and G.S.. M.P.L, R.R., S.N.C, and M.F developed the theory. S.K performed atom probe tomography and analyzed the data with B.P.W, and L.A.E.S.. M.L. fabricated heterostructure field effect transistors measured by B.P.W.. S.A. and N.S. fabricated quantum dot devices measured by B.P.W., A.J.Z., S.G.J.P., M.T.M., X.X., G.Z, and N.S.. S.N.C., O.M., M.F., and G.S. supervised the project. G.S. conceived the project. B.P.W, M.P.L, S.K, and G.S. wrote the manuscript with input from all authors.
\\
\section{Competing interests}
\noindent M.P.L., S.N.C., and M.F. declare a related patent application that proposes to intentionally introduce Ge into the quantum well, to increase the average valley splitting: US Patent Application No. 63/214957 (currently under review).

\section{Additional information}
\noindent \textbf{Supplementary Information} Supplementary Figs.~1--18, Supplementary Table~1, Supplementary Movies~1--5, and Discussion.

\noindent \textbf{Correspondence and request for materials} should be addressed to G.S.
\end{footnotesize}
\end{document}


\beginsupplement

\title{Supplementary Information: Atomic fluctuations lifting the energy degeneracy in Si/SiGe quantum dots}

\author{Brian Paquelet Wuetz}
\thanks{These authors contributed equally}
\affiliation{QuTech and Kavli Institute of Nanoscience, Delft University of Technology, PO Box 5046, 2600 GA Delft, The Netherlands}
\author{Merritt P. Losert}
\thanks{These authors contributed equally}
\affiliation{University of Wisconsin-Madison, Madison, WI 53706 USA}
\author{Sebastian Koelling}
\thanks{These authors contributed equally}
\affiliation{Department of Engineering Physics, École Polytechnique de Montréal, Montréal, Case Postale 6079, Succursale Centre-Ville, Montréal, Québec, Canada H3C 3A7}
\author{Lucas E.A. Stehouwer}
\affiliation{QuTech and Kavli Institute of Nanoscience, Delft University of Technology, PO Box 5046, 2600 GA Delft, The Netherlands}
\author{Anne-Marije J. Zwerver}
\affiliation{QuTech and Kavli Institute of Nanoscience, Delft University of Technology, PO Box 5046, 2600 GA Delft, The Netherlands}
\author{Stephan G.J. Philips}
\affiliation{QuTech and Kavli Institute of Nanoscience, Delft University of Technology, PO Box 5046, 2600 GA Delft, The Netherlands}
\author{Mateusz T. M\k{a}dzik}
\affiliation{QuTech and Kavli Institute of Nanoscience, Delft University of Technology, PO Box 5046, 2600 GA Delft, The Netherlands}
\author{Xiao Xue}
\affiliation{QuTech and Kavli Institute of Nanoscience, Delft University of Technology, PO Box 5046, 2600 GA Delft, The Netherlands}
\author{Guoji Zheng}
\affiliation{QuTech and Kavli Institute of Nanoscience, Delft University of Technology, PO Box 5046, 2600 GA Delft, The Netherlands}
\author{Mario Lodari}
\affiliation{QuTech and Kavli Institute of Nanoscience, Delft University of Technology, PO Box 5046, 2600 GA Delft, The Netherlands}
\author{Sergey V. Amitonov}
\affiliation{QuTech and Kavli Institute of Nanoscience, Delft University of Technology, PO Box 5046, 2600 GA Delft, The Netherlands}
\author{Nodar Samkharadze}
\affiliation{QuTech and Netherlands Organisation for Applied Scientific Research (TNO), Delft, The Netherlands.}
\author{Amir Sammak}
\affiliation{QuTech and Netherlands Organisation for Applied Scientific Research (TNO), Delft, The Netherlands.}
\author{Lieven M.K. Vandersypen}
\affiliation{QuTech and Kavli Institute of Nanoscience, Delft University of Technology, PO Box 5046, 2600 GA Delft, The Netherlands}
\author{Rajib Rahman}
\affiliation{University of New South Wales, Sydney, Australia\\}
\author{Susan N. Coppersmith}
\affiliation{University of New South Wales, Sydney, Australia\\}
\author{Oussama Moutanabbir}
\affiliation{Department of Engineering Physics, École Polytechnique de Montréal, Montréal, Case Postale 6079, Succursale Centre-Ville, Montréal, Québec, Canada H3C 3A7}
\author{Mark Friesen}
\affiliation{University of Wisconsin-Madison, Madison, WI 53706 USA}
\author{Giordano Scappucci}
\email{g.scappucci@tudelft.nl}
\affiliation{QuTech and Kavli Institute of Nanoscience, Delft University of Technology, PO Box 5046, 2600 GA Delft, The Netherlands}

\date{\today}
\pacs{}

\maketitle
\tableofcontents
\newpage

\section{Electrical characterization}

\subsection{Magnetotransport characterisation of Hall-bar shaped heterostructure field effect transistors}

\begin{figure*}[!ht]
	\includegraphics[width=180mm]{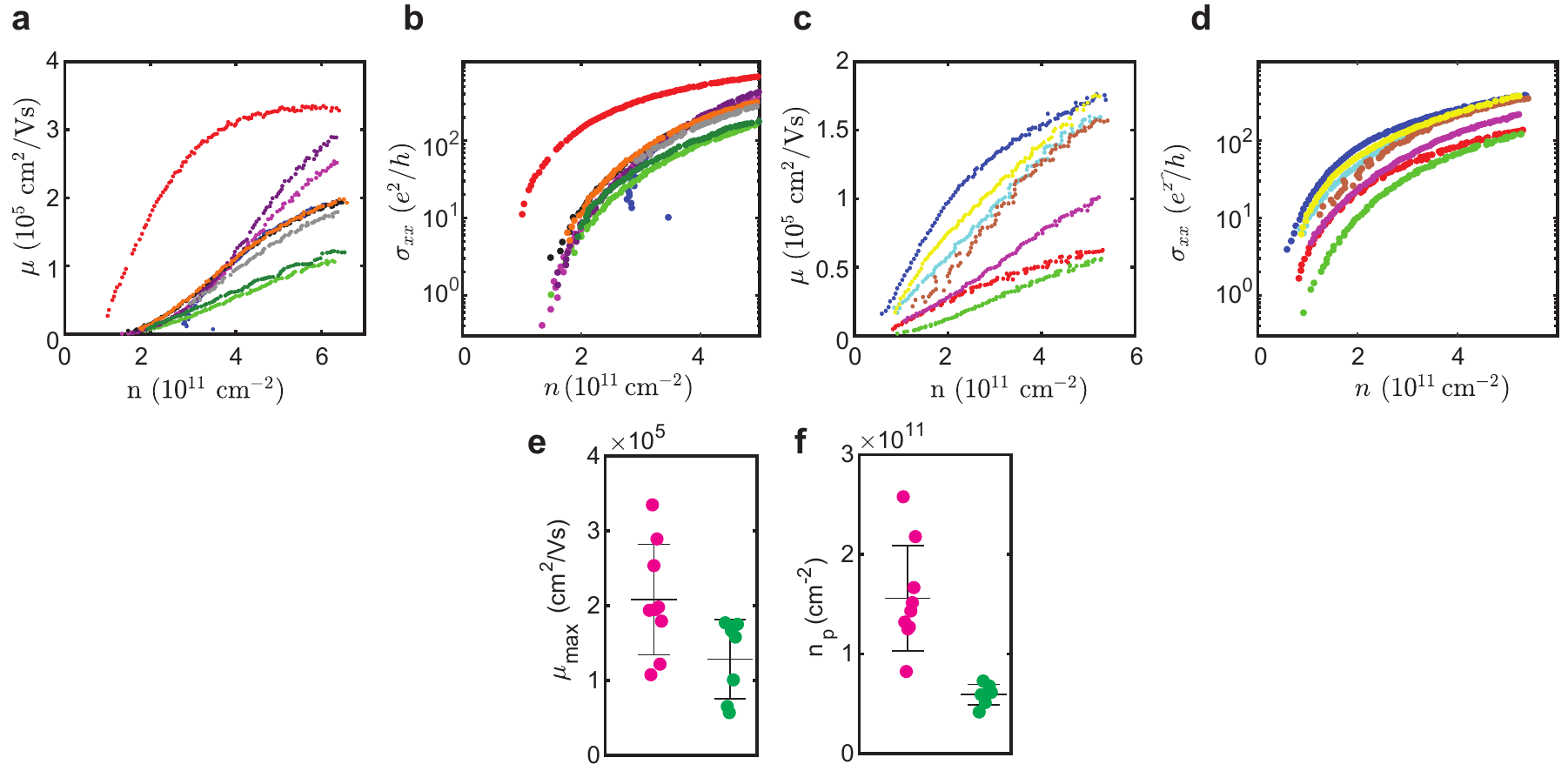}%
	\caption{\textbf{a},\textbf{b}. Mobility $\mu$ and conductivity $\sigma_{xx}$ as a function of Hall density $n$ measured for quantum well A.  \textbf{c}, \textbf{d}  Mobility $\mu$ and conductivity $\sigma_{xx}$ as a function of Hall density $n$ measured for quantum well B. \textbf{e}	Maximum mobility $\mu_{max}$ for quantum well A (magenta) and quantum well B (green) extracted from \textbf{a} and \textbf{c}. Black crosses are the mean and standard deviation. For quantum well A we find $\overline{\mu_{max}} = 129.000 \pm 53.000$ cm$^2$/Vs and for quantum well B we find $\overline{\mu_{max}} = 208.000 \pm 74.000$ cm$^2$/Vs. \textbf{f} Percolation density $n_p$ for quantum well A (magenta) and quantum well B (green) extracted by fitting the conductivity-density curves in \textbf{b} and \textbf{d} to the relationship $\sigma_{xx} \propto (n-n_p)^{1.31}$\cite{Tracy2009ObservationMOSFET}. Since this percolation theory is valid only at low densities, for each sample we chose a fitting range that goes from the lowest measured density $n_{min}$ to a density $n_{max,fit}$ that yields the best fitting results. For the devices from quantum well A in \textbf{b} we have $n_{max,fit}$ = 3.2$\times$10$^{11}$ cm$^{-2}$, 2.2$\times$10$^{11}$ cm$^{-2}$, 2$\times$10$^{11}$ cm$^{-2}$, 2$\times$10$^{11}$ cm$^{-2}$, 2.2$\times$10$^{11}$ cm$^{-2}$, 2.2$\times$10$^{11}$ cm$^{-2}$, 2.5$\times$10$^{11}$ cm$^{-2}$, 4$\times$10$^{11}$ cm$^{-2}$, 5.8$\times$10$^{11}$ cm$^{-2}$. For the devices from quantum well B in \textbf{d} we have $n_{max,fit}$ = 1.35$\times$10$^{11}$ cm$^{-2}$, 1.35$\times$10$^{11}$ cm$^{-2}$, 1.6$\times$10$^{11}$ cm$^{-2}$, 1.6$\times$10$^{11}$ cm$^{-2}$, 1.6$\times$10$^{11}$ cm$^{-2}$, 1.8$\times$10$^{11}$ cm$^{-2}$, 1.35$\times$10$^{11}$ cm$^{-2}$ 
	Black crosses are the mean and standard deviation of the percolation density. For quantum well A we find $\overline{n_{p}} = 1.56 \pm 0.53 \times$10$^{11}$ cm$^{-2}$
	and for quantum well B we find $\overline{n_{p}} = 0.59 \pm 0.1\times$10$^{11}$ cm$^{-2}$.}
\label{mobdens}
\end{figure*}

\newpage
\begin{figure*}[!ht]
\includegraphics[width=180mm]{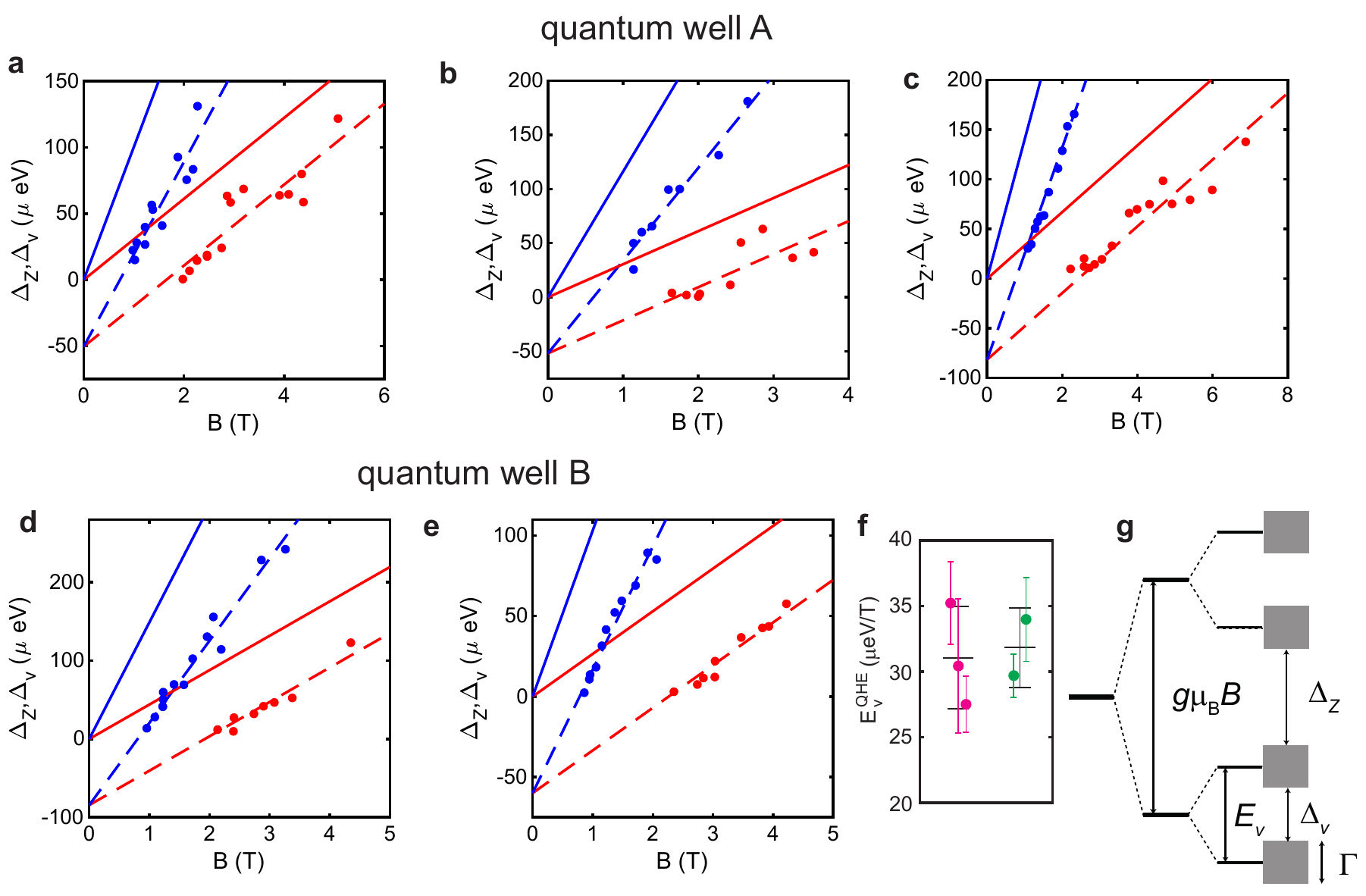}%
\caption{\textbf{a}-\textbf{c} Activation energy measurements of the valley gap $\Delta_v$ (red circles) and Zeeman gap $\Delta_Z$ (blue circles) as a function of the magnetic field $B$ for three different devices from quantum well A. $\Delta_v$ is measured at the $2n-1$ quantum Hall filling factors and $\Delta_Z$ is measured at the $4n-2$ filling factors. We follow the same methodology as in Ref.~\cite{wuetz2020effect}. The blue and red dashed lines are theoretical fits to the experimental data using the equations $\Delta_Z =g^*\mu_BB - c_BB - \Gamma$ and $\Delta_v = c_BB -\Gamma$,  where $g^*$ is the effective Land\'e-g-factor, $\mu_B$ is the Bohr magneton, $c_B$ is the proportionality factor of the valley splitting with $B$, and $\Gamma$ is the Landau level broadening induced by disorder. We obtain $c_B = 30.64\pm3.14$~\si{\micro\electronvolt}/T, $30.43\pm 5.12$~\si{\micro\electronvolt}/T, $32.46\pm 2.14$~\si{\micro\electronvolt}/T, and $g^* = 1.74\pm 0.16,\  2\pm 0.21,\  2.36\pm 0.12$ respectively. The blue and red solid lines correspond to the estimated Zeeman and valley energy gaps, respectively.	\textbf{d}, \textbf{e} Activation energy measurements and fits of the valley gap and Zeeman gap as in \textbf{a}-\textbf{c} for two devices from quantum well B. We obtain $c_B = 26.28\pm 1.65$~\si{\micro\electronvolt}/T, $43.15\pm 3.19$~\si{\micro\electronvolt}/T, and $g^* = 1.77\pm 0.13,\ 2.54\pm 0.17$ respectively. \textbf{f} Rate of increase of valley splitting with magnetic field $E_V^{QHE}$ for quantum well A (magenta) and quantum well B (green) extracted from the fitting analysis of \textbf{a}-\textbf{e}. We calculate $E_V^{QHE}$ by setting $E_V^{QHE} = c_B g/g^*$, thereby scaling $c_B$ with a coefficient $g/g^*$ that normalizes the fitted $g^*$ to the value $g=2$ in silicon. This normalization is a way to take into account the modest electron-electron interaction present in different devices, allowing for a comparison across different quantum wells. Black crosses are the mean and standard deviation of $E_V^{QHE}$. For quantum well A we find $\overline{E_V^{QHE}} = 31.1 \pm 3.9$~\si{\micro\electronvolt}/T and for quantum well B we find $\overline{E_V^{QHE}} = 31.8 \pm 3$~\si{\micro\electronvolt}/T. \textbf{g}, Schematic drawing of a Landau level split into Zeeman and valley energy levels, showing all relevant energy separations. Shaded areas represent the single-particle Landau level broadening $\Gamma$ due to disorder \cite{wuetz2020effect}.
}
\label{VS2D}
\end{figure*}

\newpage

\begin{figure*}[!ht]
	\includegraphics[width=180mm]{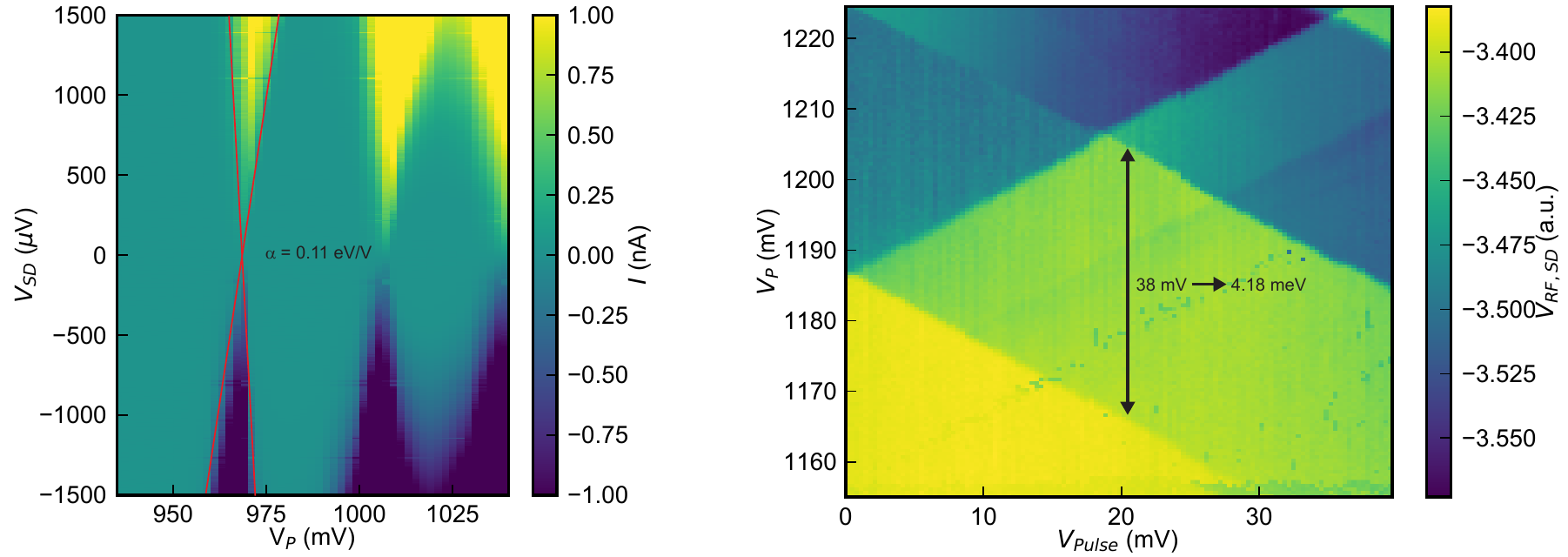}%
	\caption{\textbf{A} Coulomb blockade measurements of QD1, device 5 (see Table S3). The current through the QD is monitored while scanning the gate voltage and the bias voltage applied between the source and the drain, resulting in Coulomb diamonds. From the leftmost Coulomb diamond (indicated by the red lines) we extract a leverarm $\alpha = 0.11$~eV/V using the method described in the supplementary information of Ref.~\cite{connors2019low}. \textbf{b} Pulsed gate spectroscopy for the same quantum dot. The time-averaged RF reflectometry signal/sensing dot response is plotted as a function of the dc gate voltage $V_P$ and the square pulse amplitude $V_{pulse}$ with a pulse frequency of 25~kHz, both applied to the same gate. The arrow indicates the orbital splitting, which we extract as $E_{orb}=\alpha V_{orb}= 4.18$~meV, consitent with other values reported in literature \cite{hollmann2020large,zajac2015reconfigurable,mi2017high}.}
\label{orbital}
\end{figure*}

\newpage
\subsection{Singlet-triplet energy splitting in quantum dots}

\begin{figure*}[!ht]
	\includegraphics[width=93mm]{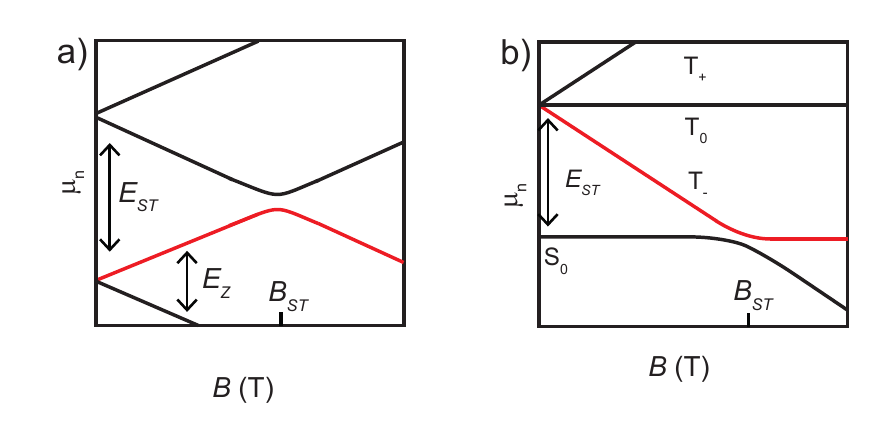}%
\caption{\textbf{a}, Energy evolution of the ground state and first excited state in a single quantum dot as a function of the magnetic field. The red line shows the expected spin filling for the charge transition $N = 1 \rightarrow 2$. 
At $B = B_{ST}$ the typical kink can be observed, where the Zeeman energy $E_Z$ is equal to the singlet-triplet splitting energy $E_{ST}$.
\textbf{b}, Energy evolution of the four lowest lying energy states in a double quantum dot as a function of the magnetic field with fixed electron number $N = 2$. The red line represents the $T_-$ energy state measured along the (1,1) $\rightarrow$(2,0) transition. At $B = B_{ST}$ the singlet state $S_0$ and the triplet state $T_-$ are equal in energy, resulting in an anticrossing.}
\label{ST}
\end{figure*}

The singlet-triplet energy splitting is computed according to the configurations in Fig.~\ref{ST}. In the configuration in Fig.~\ref{ST}a the red line can be fitted to compute $E_{ST}$ with the formula\cite{dodson2021valley}:
\begin{equation}
 	    V_P = \frac{1}{\alpha\beta_e}
 	    \ln{\frac{e^{\frac{1}{2}\kappa B+\beta_e E_{ST}}
 	    (e^{\kappa B}+1)}
 	    {e^{\kappa B}+e^{2\kappa B}
 	    +e^{\kappa B+\beta_e E_{ST}}+1}},
 	    \label{valley_12}
\end{equation}
where $\alpha$ is the lever arm converting gate voltage to energy, $V_P$ is the gate voltage, $\kappa = g\mu_B\beta_e$ where $\beta_e = 1/k_B T_e$, $g$ is the Lande-g-factor in silicon, $\mu_B$ is the Bohr magneton, $B$ is the magnetic field, $k_B$ is Boltzmann's constant, and $T_e$ is the electron temperature\cite{dodson2021valley}.

In the configuration in Fig.~\ref{ST}b the Hamiltonian of the T${-}$ state is given by:

\begin{gather}
\hat{H} = 
\begin{pmatrix}
E_{S0} & t_c\\
t_c & E_{T-}
\end{pmatrix}
\label{ref:hamT-}
\end{gather}

where $E_{S0}$ is the energy evolution of the singlet state, $E_{T-}$ is the energy evolution of the triplet minus-state, and the off-diagonal element $t_c$ is the tunnel coupling between the (1,1)-state and the the (2,0)-state in the double quantum dot. Diagonalization of the Hamiltonian yields:

\begin{equation}
 	    \mu_n(T-) = \frac{1}{2}(E_{S0}+E_{T-}+\sqrt{(E_{S0}-E_{T-})^2+4t_c^2} \ )
 	    \label{valley_T-}
\end{equation}
To fit the red line from Fig.~\ref{ST}b we use $E_{S0} = 0$ and $E_{T_-} = \alpha(g\mu_B B+E_{ST})$, where $\alpha$ is the lever arm, $g$ is the single particle g-factor, $B$ is the magnetic field, and $E_{ST}$ is the singlet-triplet splitting.

\newpage
\begin{figure*}[!htp]
	\includegraphics[width=180mm]{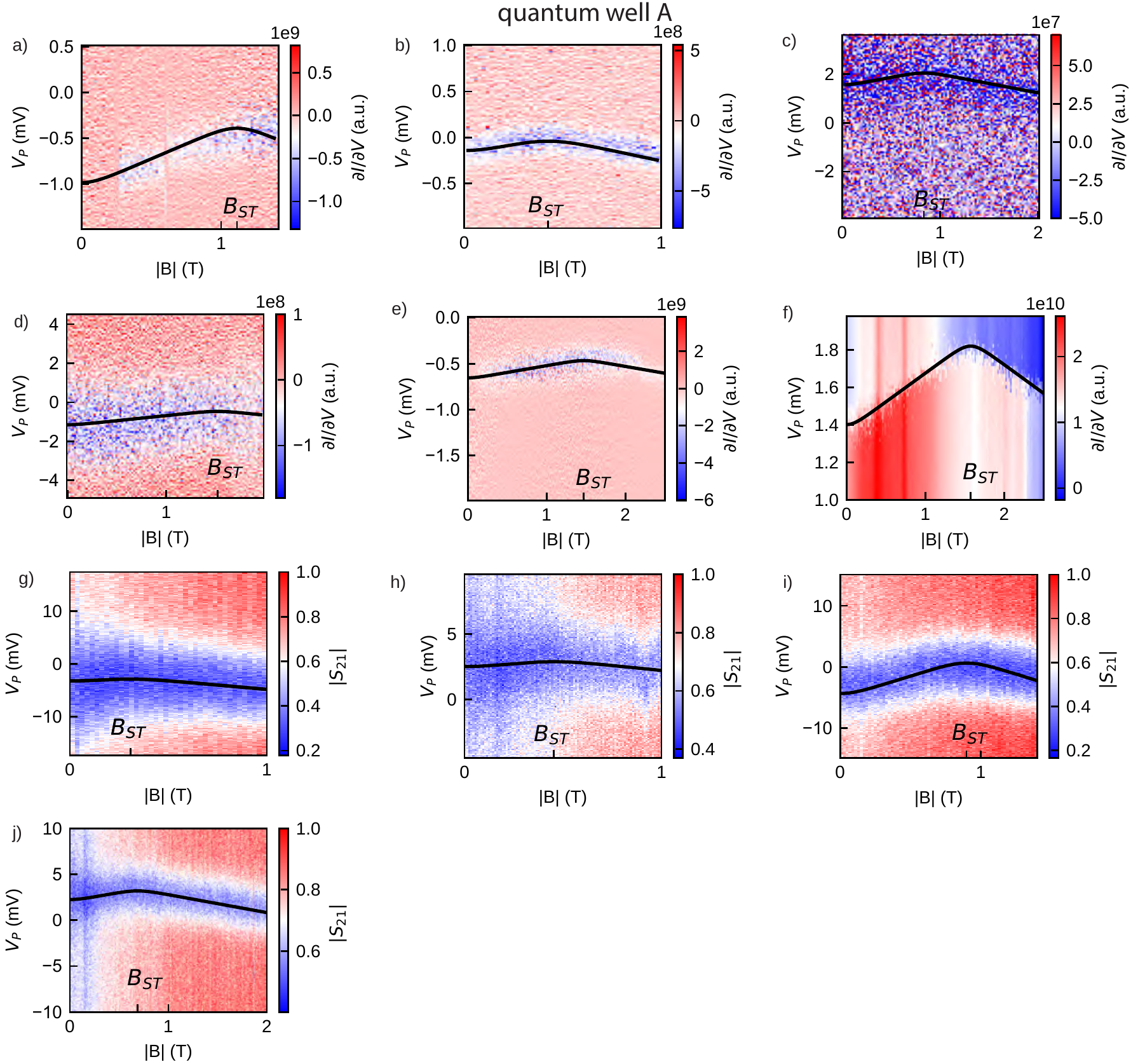}%
	\caption{Magnetospectroscopy of quantum dots fabricated on quantum well A. $V_P$ is the gate voltage applied to the plunger gate forming the quantum dot. For clarity, we subtract from $V_P$ in panels a) - j) an offset that depends is on the quantum dot being measured. a) - f) Magnetospectroscopy data measured along the $N=1\rightarrow2$ transition of five different quantum dots on three different samples in quantum well A. The signal is measured by monitoring the derivative of the current through a nearby charge sensor. a), A charge fluctuation occurred during the measurement and to optimize the fitting routine, we shifted the data in the range 0.3-0.6 T upwards by 1 mV. a) - f), Due to low tunnel rates, for each gate Voltage sweep at the different magnetic fields, we determine the points with the highest derivative of the current $\frac{\partial I}{\partial V}$ through the charge sensor as the $N=1\rightarrow2$ charge transition. We then use these points as the input of eq. \ref{valley_12}. With this equation we can fit the charge transition as a function of the magnetic field (black curve). g) - j) Magnetospectroscopy data measured along the $N=1\rightarrow2$ transition of four different quantum dots on two different samples in quantum well A. The quantum dot is probed via gate-based sensing using an on-chip superconducting resonator in these measurements \cite{samkharadze2018strong}. The magnitude of the transmitted microwave signal S$_{21}$ through a feed line that is capacitively coupled to the resonator is plotted here. For each gate Voltage sweep at the different magnetic fields, we use a Lorentzian function to find the resonance peak of the signal. The resonance peaks then are used as input of eq. \ref{valley_12}. With this equation we can fit the charge transition as a function of the magnetic field (black curve).}
\label{stackA}
\end{figure*}
\newpage

\begin{figure*}[!htp]
	\includegraphics[width=180mm]{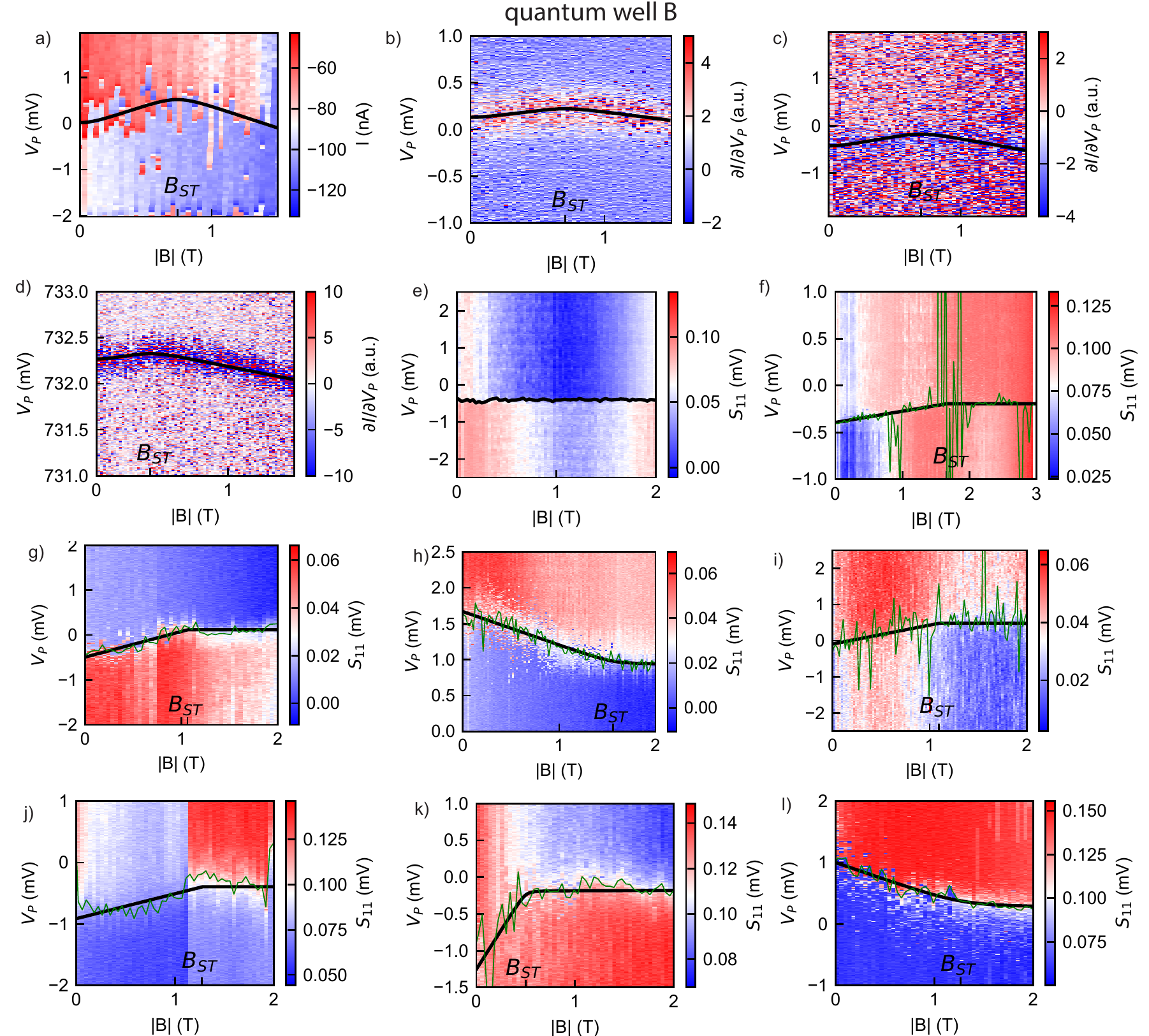}%
	\caption{Magnetospectroscopy of quantum dots fabricated on quantum well B. $V_P$ is the gate voltage applied to the plunger gate forming the quantum dot. For clarity, we subtract from $V_P$ in panels a) - l) an offset that depends is on the quantum dot being measured. a) -d) Magnetospectroscopy data measured along the $N=1\rightarrow2$ transition of four different quantum dots on two different samples in quantum well B. The signal is measured by monitoring \textbf{a}, the current $I$ through a nearby charge sensor, or \textbf{b} - \textbf{d} by monitoring the derivative of the current $\frac{\partial I}{\partial V}$ through a nearby charge sensor. \textbf{a} - \textbf{d}, To extract the inflection point of the electron charge transition, we fit the signal of the detuning for every magnetic field to eq. (2) from Ref. \cite{PhysRevLett.92.226801}. The inflection points then are used as input of eq. \ref{valley_12}. With this equation we can fit the charge transition as a function of the magnetic field (black curve). e) - l) Magnetospectroscopy data measured along the $N=(1,1)\rightarrow(2,0)$ transition of eight different quantum dots on two different samples in quantum well B. The signal is measured by monitoring the reflected amplitude of the rf readout signal through a nearby charge sensor. To extract the inflection point of the electron charge transition, we fit the signal of the detuning for every magnetic field to eq. (2) from Ref. \cite{PhysRevLett.92.226801}. Here we superimpose the inflection points as green curves, to help the reader to follow the charge transitions. To extract $B_{ST}$ we use the crossing point of two linear fits (black solid lines) along the $T_-$ and $S_0$-state. On top of these samples there is a micromagnet lowering the magnetic field strength at the center of the sample by up to 0.2~T corresponding to 23~\text{$\mu$}eV which is taken as a lower bound for measurable $E_{ST}$.}
\label{stackB}
\end{figure*}
\newpage

\begin{table}[!ht]
\begin{tabular}{|l|l|l|l|l|l|l|l|l|}
\hline
Stack & Wafer ID & database processing ID & Figure & device ID&transition & $B_\text{ST}$~(T) & $E_\text{ST}$~($\si{\micro\electronvolt})$ &$d_p$~(nm)\\  \hline
QW A&QT428 & DEMO 13& S4a &D1 2-dot, P2 & (0,1) $\rightarrow$ (0,2)  & 1.11 & 129$\pm$1.1 & 50 \\ \hline
QW A&QT428  &DEMO 13 & S4b &D1 2-dot, P1 & (0,1) $\rightarrow$ (0,2) & 0.42 & 49.4$\pm$2.2 & 50 \\ \hline
QW A &QT428 &DEMO 21 &S4c &D2 2-dot, P1 & (0,1) $\rightarrow$ (0,2) & 0.83 &  96.6$\pm$6.3 & 50 \\ \hline
QW A &QT428 &DEMO 21 &S4d &D2 2-dot, P2 & (0,1) $\rightarrow$ (0,2) & 1.47 & 170.4$\pm$9.0 & 50 \\ \hline
QW A &QT428 &DEMO 15 &S4e &D3 2-dot, P1 & (0,1) $\rightarrow$ (0,2) & 1.52 & 176.3$\pm$13.4 & 50 \\ \hline
QW A&QT428 &DEMO 15 &1f, S4f &D3 2-dot, P2 & (0,1) $\rightarrow$ (0,2)  & 1.57 & 182.3$\pm$5.8 & 50 \\ \hline
QW A &QT539 & SQ19-193-1-3-03&S4,  g&D4 2-dot, P1 & (0,1) $\rightarrow$ (0,2) & 0.31 & 35.7$\pm$5.9 & 50  \\ \hline
QW A &QT539  & SQ19-193-1-3-03&S4, h &D4 2-dot, P2 & (0,1) $\rightarrow$ (0,2) & 0.45 & 52.6$\pm$0.8 & 50  \\ \hline
QW A &QT539  & SQ19-193-1-3-04&S4, i &D5 2-dot, P1 & (0,1) $\rightarrow$ (0,2) & 0.9 & 104$\pm$1.6 & 50  \\ \hline
QW A &QT539  & SQ19-193-1-3-04&S4, j &D5 2-dot, P2 & (0,1) $\rightarrow$ (0,2) & 0.69 & 79.6$\pm$2.0 & 50 \\ \hline
QW B & QT592 &SQ20-20-5-25-2 &S5, a &D1 5-dot, P4 & (0,1) $\rightarrow$ (0,2) & 0.74 & 85.7$\pm$2.0 & 40 \\ \hline
QW B  & QT592 &SQ20-20-5-25-2 &S5, b &D1 5-dot, P1 & (0,1) $\rightarrow$ (0,2) & 0.71 &  82.1$\pm$3.7 & 40 \\ \hline
QW B  &QT592 &SQ20-20-5-25-2 &S5, c &D1 5-dot, P2 & (0,1) $\rightarrow$ (0,2) & 0.7 & 81.7$\pm$10.1 & 40  \\ \hline
QW B  & QT553 & SQ19-228-2-44-2 &S5, d &D6 2-dot, P2 & (0,1) $\rightarrow$ (0,2) & 0.41 & 47.2$\pm$3.68 & 50 \\ \hline
QW B  &QT592 & SQ20-20-5-18-4 &S5, e &D1 6-dot, P3 & (1,1) $\rightarrow$ (0,2) & 0 & 0$\pm$0 & 50 \\ \hline
QW B  &QT592 & SQ20-20-5-18-4 &S5, f &D1 6-dot, P4 &  (1,1) $\rightarrow$ (0,2) & 1.73 & 191.5$\pm$13.2 & 50 \\ \hline
QW B  &QT637 & SQ20-205-2-12 &S5, g &D2 6-dot, P1 &  (1,1) $\rightarrow$ (0,2) & 1.06 & 123.1$\pm$8.9 & 40 \\ \hline
QW B  &QT637 &SQ20-205-2-12 &S5, h &D2 6-dot, P2 & f (1,1) $\rightarrow$ (0,2) & 1.56 & 180.5$\pm$9.7 & 40 \\ \hline
QW B  &QT637 &SQ20-205-2-12 &S5, i &D2 6-dot, P3 &  (1,1) $\rightarrow$ (0,2) & 1.1 & 126.8$\pm$33.6 & 40 \\ \hline
QW B  &QT637 &SQ20-205-2-12 &S5, j &D2 6-dot, P4 &  (1,1) $\rightarrow$ (0,2) & 1.27 & 147.3$\pm$15.7 & 40 \\ \hline
QW B  &QT637 &SQ20-205-2-12 &S5, k &D2 6-dot, P5 &  (1,1) $\rightarrow$ (0,2) & 0.5 & 57.9$\pm$13.5 & 40 \\ \hline
QW B  &QT637 &SQ20-205-2-12 &S5, l &D2 6-dot, P6 &  (1,1) $\rightarrow$ (0,2) & 1.25 & 144.6$\pm$19.1 & 40 \\ \hline
\end{tabular}
\caption{Summary of quantum dot valley splitting measurements. Among all devices measured, in one case (data point $E_\text{ST}=0$~$\si{\micro\electronvolt}$) we did not observe in magnetospectroscopy the signature kink associated with valley splitting. This indicates a very small valley splitting, below the lower bound of about 23~$\si{\micro\electronvolt}$ set by our experimental measurement conditions}. While very small valley splitting values are within the predicted theoretical distributions in the main text, previous theories\cite{Friesen2007ValleyWells} suggest that they could also originate from the presence of an atomic step within the quantum dot.
\end{table}

\newpage

\section{Material characterization}

\subsection{Atom Probe Tomography analysis of interfaces}
Atom Probe analysis (APT) of the interfaces is done in 5 steps. All of them explained in detail below.
First, the entire measurement is reconstructed using the standard reconstruction algorithms \cite{bas_general_1995}.
Second, a cube approximately representing the size of an electrical defined quantum in the x,y-plane and comfortably comprising the entire quantum well in the z-direction/depth-direction is extracted from the reconstructed data. This is done to have comparable sizes for each measurement, to limit the known reconstruction artefacts of APT \cite{rolland_new_2017} and to enable a direct comparison to simulations in step 5.
Third, the three-dimensional point cloud created in the usual APT reconstruction \cite{bas_general_1995} is tessellated using a Voronoi tessellation \cite{Voronoi+1908+97+102, Voronoi+1908+198+287}. The Voronoi tessellation is used for all subsequent steps. It can be viewed as a smoothing operation that “spreads out” the detected ions/atoms to a finite volume rather than representing them as zero-dimensional points.
Forth, a x,y-grid is defined on the cube and for each cell of the grid a profile based on the Voronoi tessellation along the z-axis is created that is than fitted with a sigmoid function. The collection of sigmoid functions is then used to represent the interface and calculate the interface positions as well as the isoconcentration surfaces.
Fifth, the profile extracted from the Voronoi grid of the entire cube is used to create a model structure with the known crystal structure of SiGe and a pseudo-random distribution of Si and Ge atoms in the x-y plane, enforcing the same profile along the depths direction as given by the Voronoi grid and the same percentage of atoms in the volume as expected from the detection efficiency of the Atom Probe (here: 80~\% detection efficiency of the LEAP 5000XS). These model structures interface are then compared to the measurement results.
All data treatment is done in Python 3.9 using numpy 1.20.3 and scipy 1.6.3.

\subsection{Extraction of the cubes and Voronoi tessellation}
\begin{figure*}[!ht]
	\includegraphics[width=180mm]{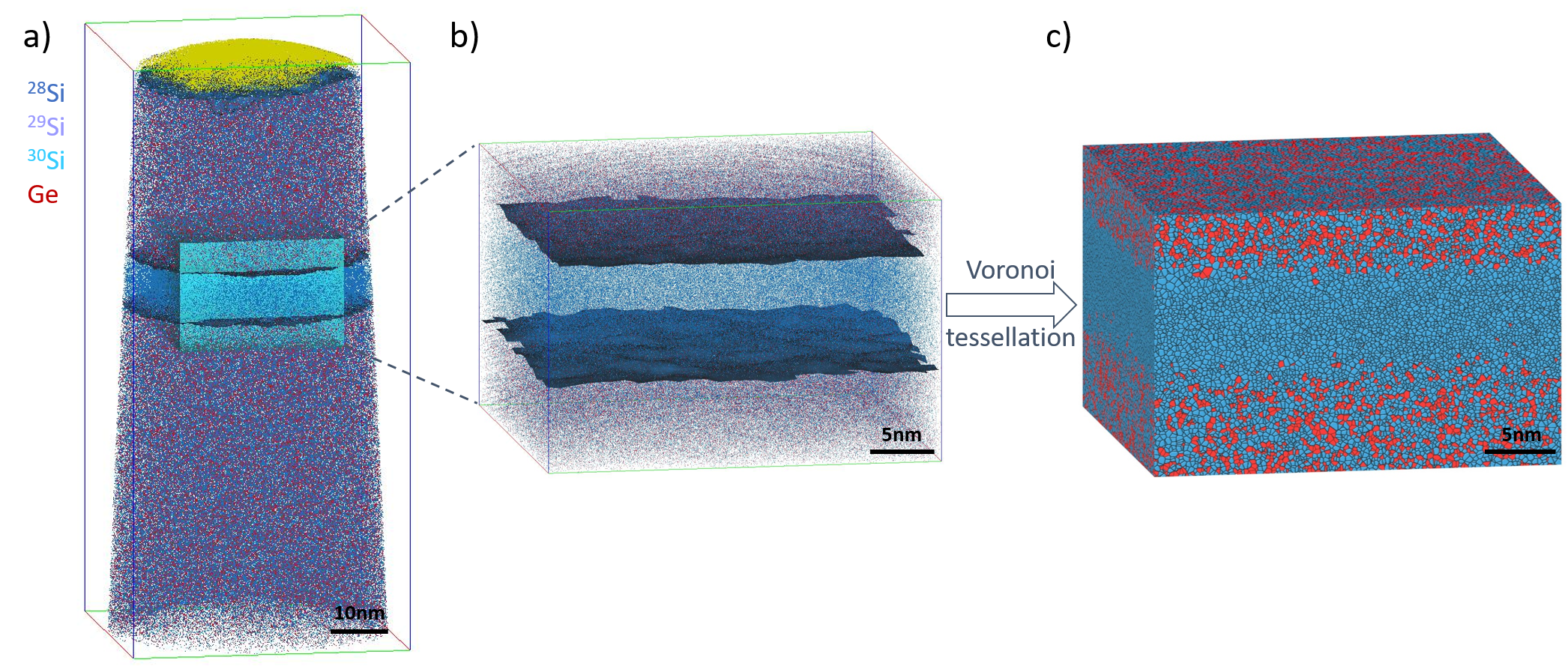}%
	\caption{Visualization of the extraction (a) of the cube (b) from the full data set (a) and Voronoi tessellation of the cube (c)}
\label{voronoi}
\end{figure*}

The cubes are manually extracted from the reconstructed volume as exemplary shown in Fig.~\ref{voronoi} a-b).
After a cube containing the quantum well with the approximate size of an electrically defined quantum dot ($\sim$ 30x30x20~nm) is extracted a Voronoi tessellation is performed on the point cloud representing APT data inside the cube. A result of such a tessellation in exemplary shown in Fig.~\ref{voronoi} c).

\subsection{Construction of the interface}
\begin{figure*}[!ht]
	\includegraphics[width=180mm]{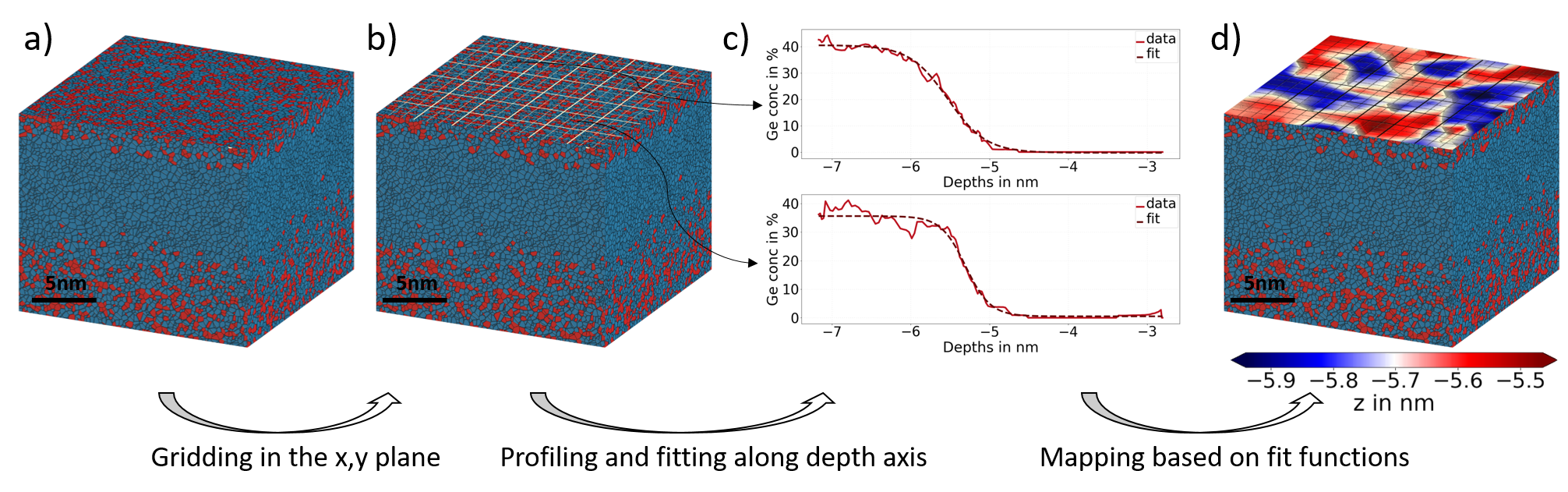}%
	\caption{Creation of a map from the Voronoi tessellated cube (a) by applying an x,y-grid (b) and fitting of profiles along z-axis with a sigmoid function in each cell (c). The profiles can then be used to calculate the position of e.g. the 25~\% Germanium isoconcentration surface (d).}
\label{sigmoidmap}
\end{figure*}

Interfaces are constructed based on the Voronoi tessellated data sets. The process is depicted in Fig.~\ref{sigmoidmap}. A grid is created in the x,y-plane of the tessellated data set (Fig.~\ref{sigmoidmap} a-b). For each cell of the grid a one-dimensional profile along the z-axis is generated using the tessellation.
As opposed to “regular” APT data \cite{bas_general_1995} where profiles are created utilizing small bins along the z-axis and concentrations are then calculated from the ions/atoms within the bin \cite{larson_local_2013} Chapter 7, the profiles on the tessellated data are created by a set of cutting planes. The process works by cutting the tessellation at each depth and use every ion/atom whose volume is cut as part of the plane and hence have it contribute to the concentration measured within that plane and at that depth.This can be viewed as a smoothing operation that spreads out the detected ions/atoms to a finite volume.

\begin{figure*}[!hp]
	\includegraphics[width=180mm]{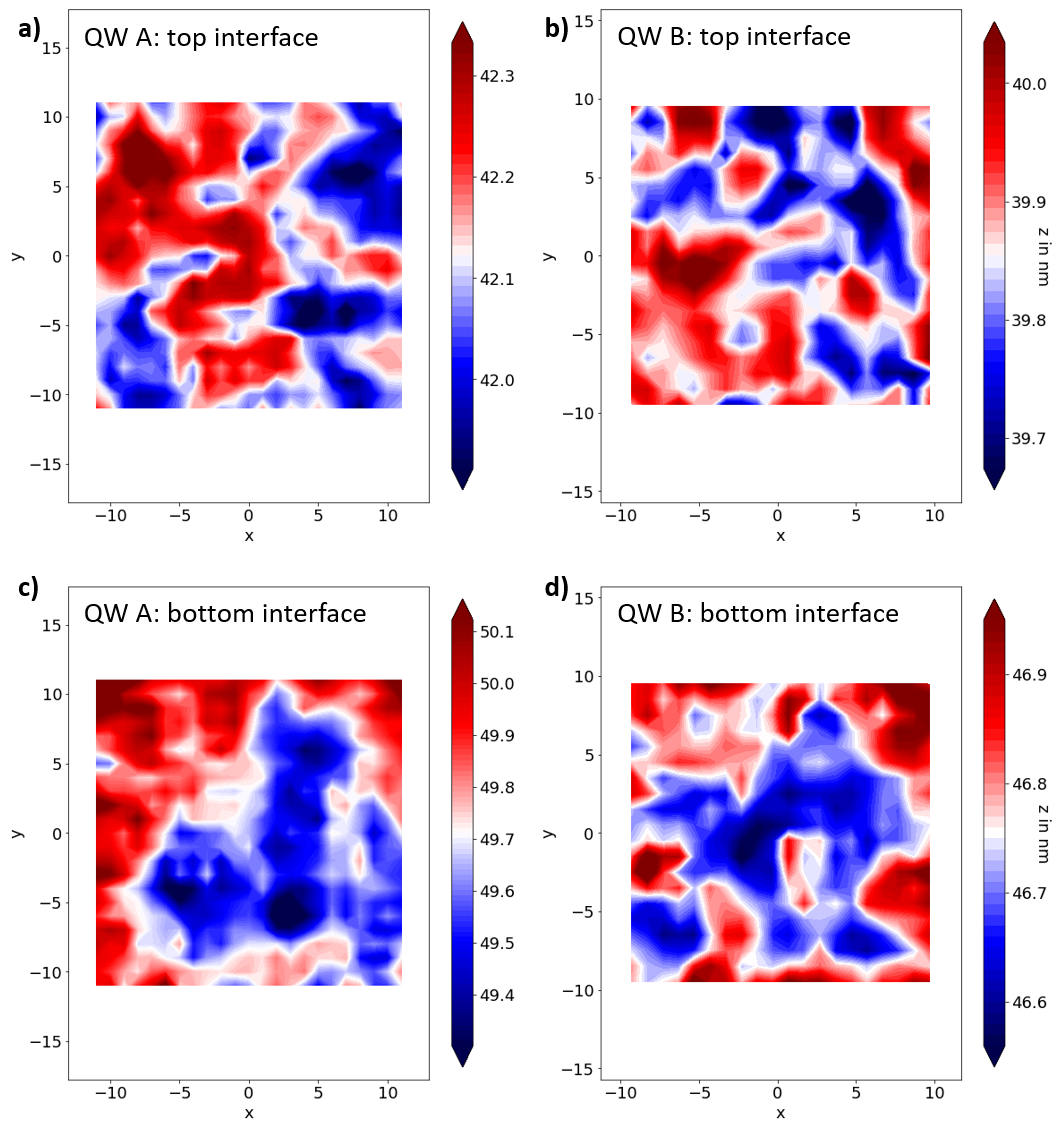}%
	\caption{Examples of position maps of top (a, b) and bottom (c, d) Germanium interfaces for both Quantum wells A and B. For each cell the depth plotted on the map is extracted from inflection point the sigmoid fit to the profile extracted from the cell (Fig.~\ref{sigmoidmap} b-c).}
\label{posmaps}
\end{figure*}

\begin{figure*}[!hp]
	\includegraphics[width=180mm]{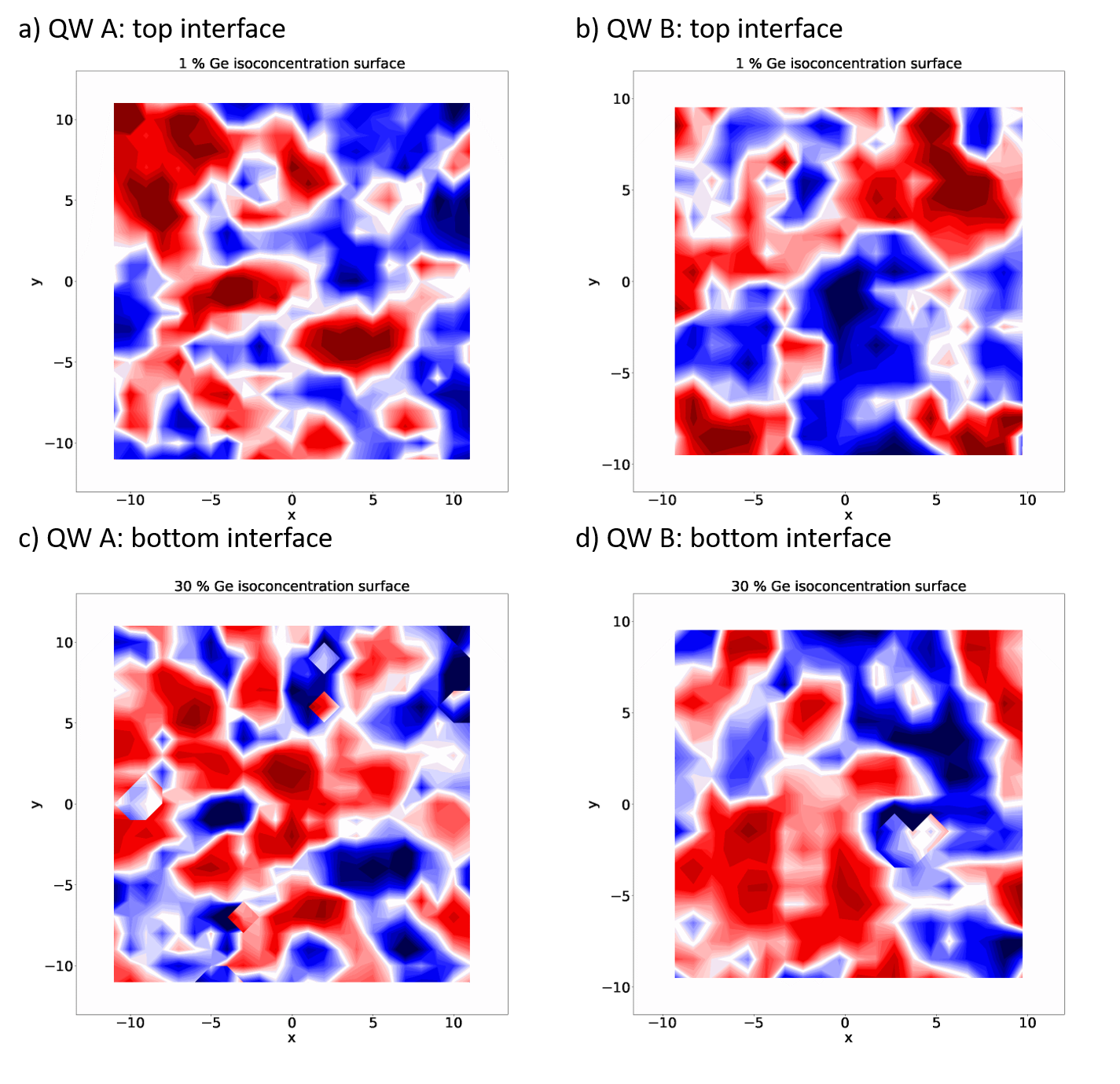}%
	\caption{Example of Germanium isoconcentration surfaces or the top (a, b) and bottom (c, d) interfaces of both Quantum Wells a and B. The plots reported here show one particular isosurface, 1\% in a, b and 30\% in c,d. As before the depth for each map can be extracted from the sigmoid fits to the profile in each cell.}
\label{isomaps}
\end{figure*}

Each x,y-cell (typically 3x3~nm wide spaced 1~nm apart and hence partially overlapping) generates a profile and is then fitted using sigmoid function \cite{dyck2017accurate} as shown in Fig.~\ref{sigmoidmap} c).
The sigmoid functions are then used to represent the interface in the following way:
\begin{itemize}
\item The inflection point of the sigmoid represents the position of the interface in each cell (Fig.~\ref{sigmoidmap} d)
\item Isoconcentration surfaces \cite{larson_local_2013} Chapter 6.3.2 are created by plotting the position where the sigmoid of each cell reaches the respective concentration
\end{itemize}
Fig.~\ref{posmaps} and \ref{isomaps} show examples of the interface positions maps and isoconcentration surface maps generated in this way for the top and bottom interfaces of a QW A and a QW B sample. Note, that the data can now readily be used to calculate the average roughness and root mean square roughness in the usual way \cite{gwyddion_rms}.

\clearpage

\subsection{Generating model data}
Model data are generated based on the known crystal properties of Si$_{66.5}$Ge$_{33.5}$. A crystal of the same size as the cubes extracted from the data ($\sim$ 30x30x20~nm) is generated digitally and then 20~\% of the atoms in the crystal are pseudo-randomly removed to account for the detection efficiency of the LEAP5000XS system used in the APT analysis.

\begin{figure*}[!ht]
	\includegraphics[width=180mm]{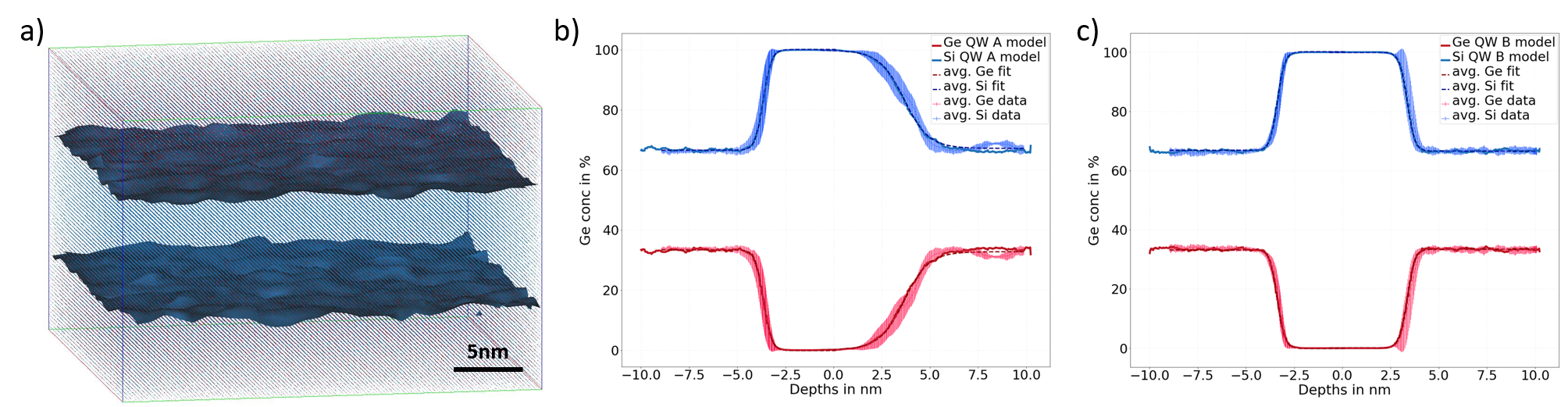}%
	\caption{Example of a crystalline cube of QW A (a) and a comparison of the average profiles of the measured quantum wells (see Fig. 2c of the main text) and profiles from a generated cube of Quantum Well A (b) and Quantum Well C (c).}
\label{modelgen}
\end{figure*}

\begin{figure*}[!hp]
	\includegraphics[width=180mm]{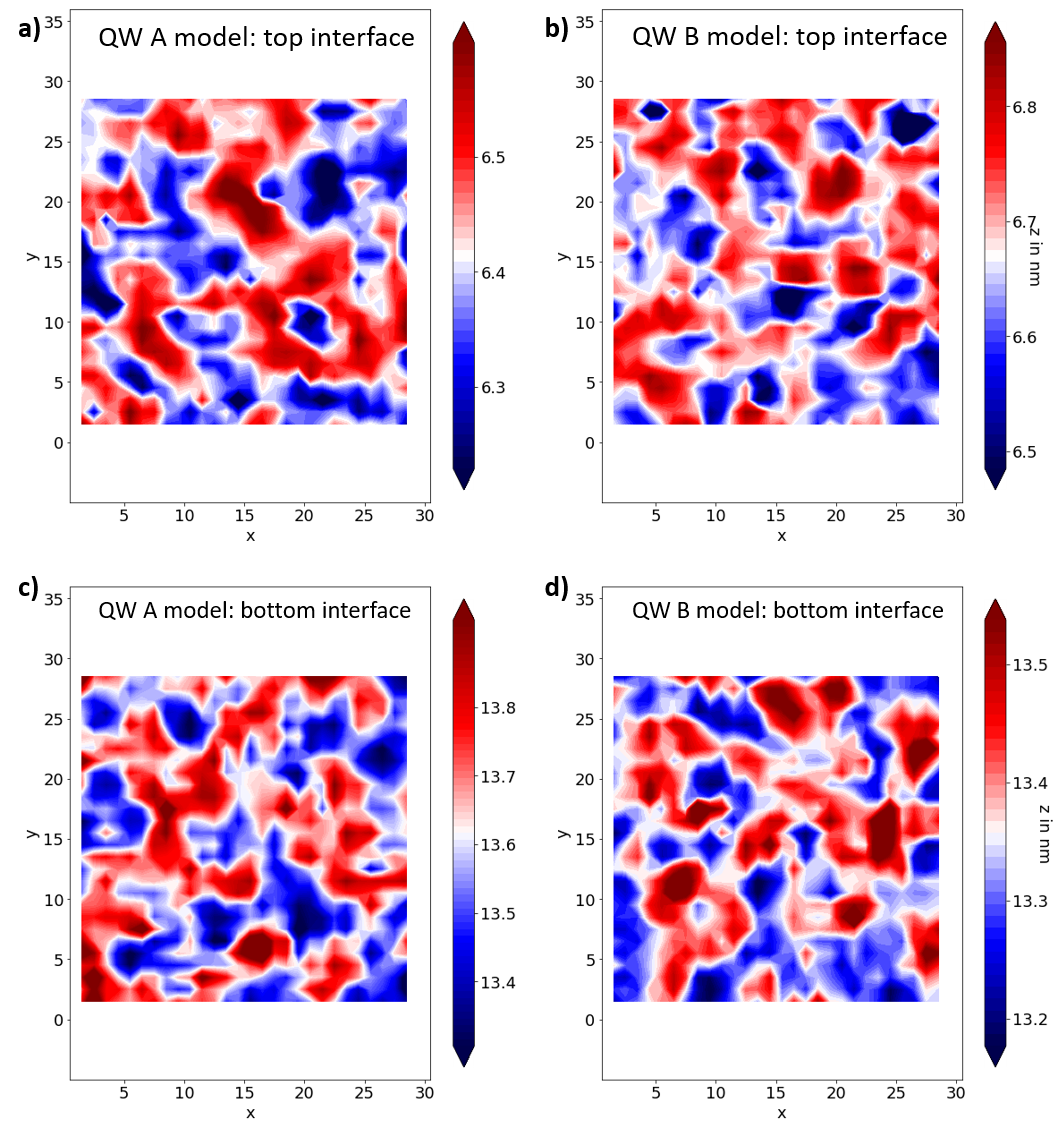}%
	\caption{Examples of position maps of top (a, b) and bottom (c, d) Germanium interfaces for model data sets of both Quantum wells A and B. As in Fig.~\ref{posmaps} the depth plotted on the map is extracted from the inflection point of the sigmoid fit for the profile along the depth axis generated in each cell (Fig.~\ref{sigmoidmap} b-c).}
\label{modelmaps}
\end{figure*}

Along the depth axis of the cube the average measured APT profile of the Si and Ge concentration of QW A and QW B as shown in Fig. 2c) of the main text is enforced. The result of the generation of such a cube for QW A and the comparison of the depth profile extracted from a cube of QWA and QW B to the average profile of QW A and QW B respectively are shown in Fig.~\ref{modelgen}.
In Fig.~\ref{modelmaps} interface position maps of these model structures are shown. They should be compared to Fig.~\ref{posmaps} where the same maps are extracted from measured data sets. The root mean square roughness as measured from the model is compared to the data measured from the APT data in Fig 2 of the main text.

Note: the animation in the file Supplementary\_ Movie\_1.m4v shows for the top interface of quantum well B  (for increasing Ge concentration) the deviation of each isosurface tile position from the isosurface’s average position. Here we benchmark the experimental data from our APT analysis (at each frame of the animation) against average and min-max range covered by 100 random models.

\clearpage

\subsection{Atomic steps, Quantum well width, and bottom interfaces}

\begin{figure*}[!ht]
	\includegraphics[width=165mm]{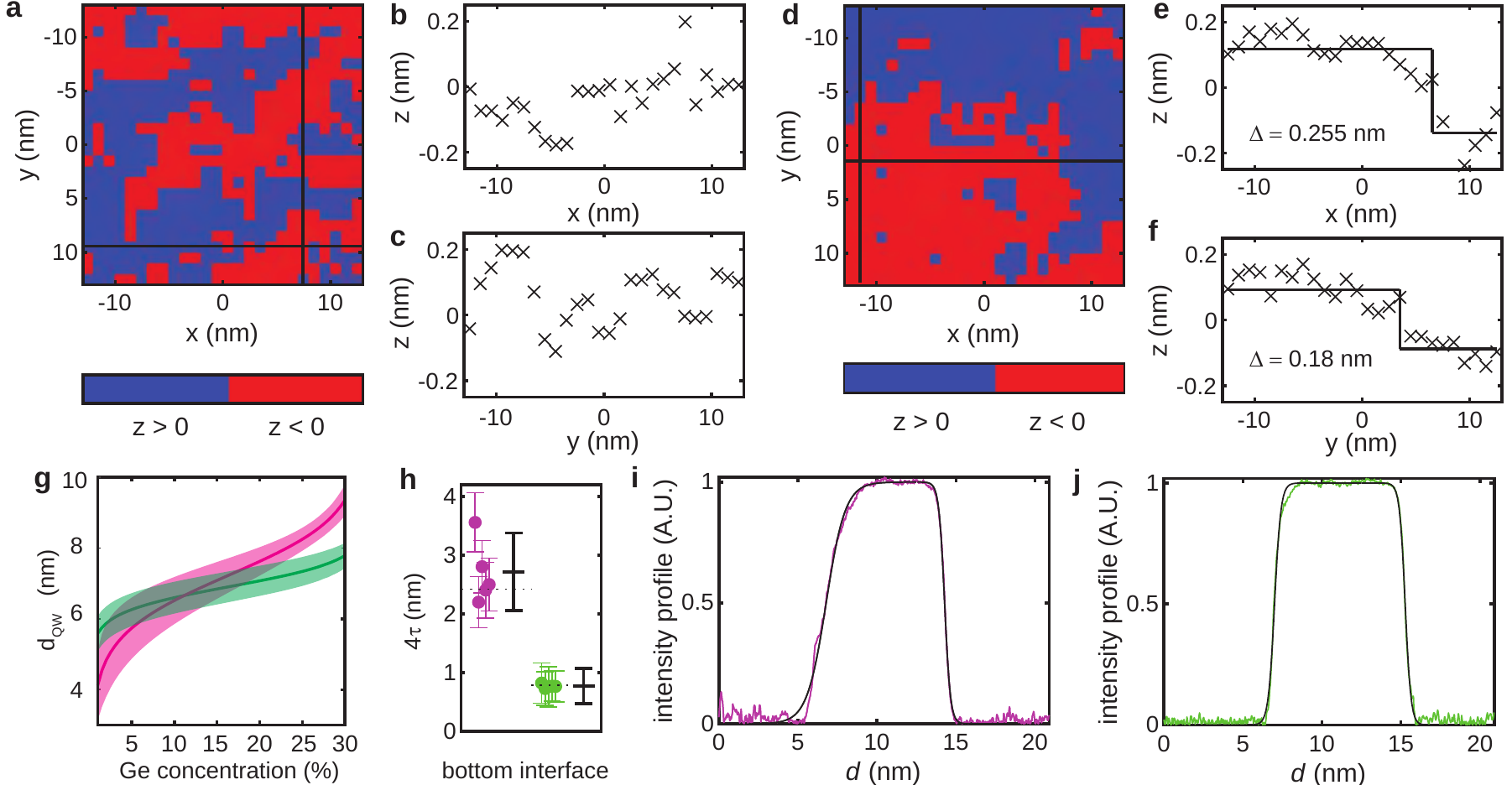}
	\caption{\textbf{a} 10\% isoconcentration surface from a Stack A sample without step. Blue areas are below and red areas above the average height (defined as $z = 0$) of the isoconcentration surface. The black lines are the positions of the line cuts in \textbf{b} and \textbf{c}. \textbf{b} line cut along the x direction of the isoconcentration surface. in \textbf{a}. \textbf{c} line cut along the y-direction of the isoconcentration surface. The z-position randomly oscillates around the mean value. \textbf{d} 10\% isoconcentration surface from a Stack A sample with clear spatial division of the blue and red areas. The black lines are the positions of the line cuts in \textbf{e} and \textbf{f}. \textbf{e} line cut along the x-direction of \textbf{d}. A step with height $\Delta = 0.255$~nm occurs at $x = 7$~nm, corresponding to approximately 2 monoatomic layers. The black line represents the Heavyside step function with the highest $C$ and the step height is determined by taking average z-position of the line cut before and after the step. \textbf{f} line cut along the y-direction of \textbf{c}. A step with $\Delta = 0.18$~nm occurs at $x = 3$~nm, corresponding to approximately 1.5 monoatomic layers. The black line represents the average z-position before and after the step. \textbf{g}, Average width of quantum well A (magenta line) and B (green line) as a function of the Ge concentration of the isoconcentration surfaces. Shaded areas represent the standard deviation of the quantum wells. \textbf{h}, Statistical analysis of the bottom $4 \tau$ interface widths derived from the fitting the data for quantum well A (magenta) and quantum well B (green). Black crosses are the mean and standard deviation for data from the different APT samples, highlighting the uniformity of the interfaces. \textbf{i}, \textbf{j}, HAADF-STEM intensity profile for stack A and B (magenta and green line, respectively) along the heterostructure growth direction (see TEMs in the main section). The black lines are fits of the data in the interface regions, using a sigmoid function.
	}
\label{step}
\end{figure*}

To evaluate the presence of atomic steps from isoconcentration surfaces, we consider one-dimensional line cuts along the x- and y-axis of an isosurface. If a line cut crosses an atomic step along the isosurface, the line cut should resemble a Heavyside step function $H$:
	\begin{equation}
     H(x-x_s) = h_0 + \begin{cases}
    -a/2, & \text{for $x<x_s$}.\\
    a/2, & \text{for $x \geq x_s$}.
        \end{cases}
	\end{equation}
	where $a$ is the step height, $x_s$ is the step position and the offset $h_0$. To quantify the resemblance between a line cut and the step function, we determine the correlation coefficient $C$ between the two with:
	\begin{equation}
	    C = \frac{\sum _k(z_k-\Bar{z})(h_k-\Bar{h})}{ \sqrt{\sum_k(z_k-\Bar{z})^2\vphantom{\sum_k(h_k-\Bar{h})^2}} \sqrt{\sum_k(h_k-\Bar{h})^2}}  
	\end{equation}
	where $z_k$ are the z-values of the line cut, $\Bar{z}$ is the mean value of the line cut, k is the index of the, $h_k$ are the values of the step function, and $\Bar{h}$ the mean value of the step function. If $C \geq 0.75$ we consider the linecut to represent a step. We subsequently can determine $a$ by taking the difference between the two plateaux $\Delta = \Bar{z}_{k+} - \Bar{z}_{k-}$, where $\Bar{z}_{k+}$ and $\Bar{z}_{k-}$ are the average z-position before and after $x_s$, respectively.\\

\newpage
\subsection{SIMS and crosshatch pattern}

\begin{figure*}[!ht]
	\includegraphics[width=183mm]{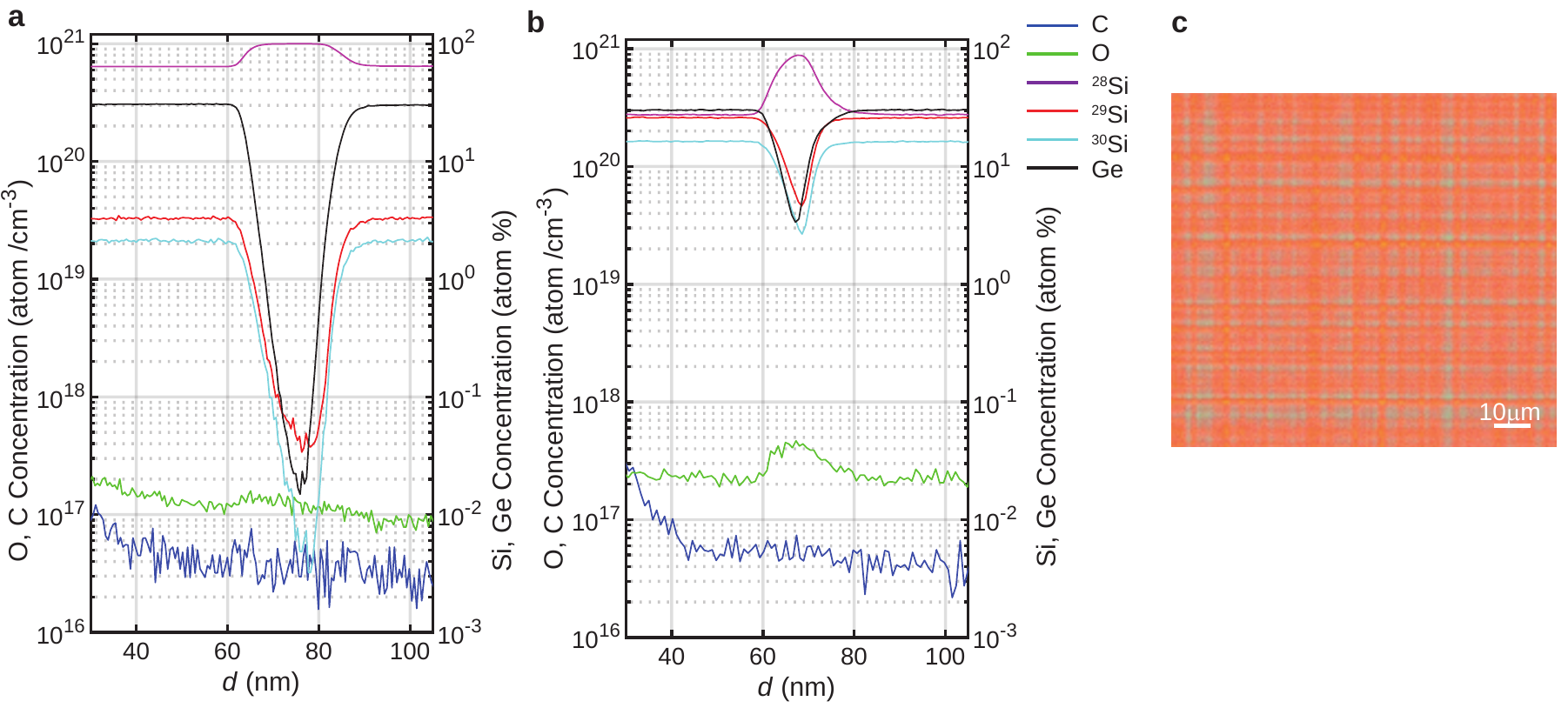}%
	\caption{\textbf{a}, \textbf{b}, Depth concentration SIMS profile of quantum well A and quantum well B respectively. Analyzed elements are $^{28}$Si (red), $^{29}$Si (blue), $^{30}$Si (purple), Ge (black), oxygen (green) and carbon (blue). In quantum well A both carbon and oxygen concentrations are below their respective detection limits of $3 \times 10^{16}$~cm$^{-3}$ and $1 \times 10^{17}$~cm$^{-3}$. In quantum well A only carbon is below the detection limits, while there is a residual oxygen content of $4 \times 10^{17}$~cm$^{-3}$ in the quantum well. \textbf{c} typical cross-hatch pattern from the surface of the wafers.}
\end{figure*}
\newpage

\section{Theoretical model}

All numerical simulations and analysis in this section were performed in Python, using the open-source libraries NumPy, SciPy, and Matplotlib.

\subsection{Tight-binding model}

In our analysis, we use the one-dimensional two-band tight-binding model of Boykin {\it et al.}~\cite{Boykin:2004:p165325}. Here, the nearest-neighbor and next-nearest-neighbor hopping amplitudes, chosen to yield the correct valley minimum wavevector and out-of-plane effective mass, are $v = 0.683$ and $u = 0.612$, respectively. The onsite energy $\varepsilon$ is the sum of the contributions from the electrostatic potential energy $eEz$, for a vertical electric field $E = 0.0125$~V/nm, and the quantum well confinement potential  $U(z)$. The conduction band offset is taken to be a linear interpolation:
\begin{equation}
    \Delta E_c = \left(x_w - x_s \right) \left[ {\frac{x_w}{1 - x_s}} \Delta E_{\Delta_2}^\mathrm{Si}(x_s) - {\frac{1 - x_w}{x_s}} \Delta E_{\Delta_2}^\mathrm{Ge}(x_s) \right] ,
\end{equation}
where $x_w$ is the Si concentration inside the well and $x_s$ is the Si concentration in the quantum well barriers (the substrate). The functions $\Delta E_{\Delta_2}^\mathrm{Si (Ge) }(x)$ describe the $\Delta_2$ conduction band offsets for strained Si (Ge) grown on an unstrained $\mathrm{Si}_x\mathrm{Ge}_{1-x}$ substrate. 
These functions are approximately linear in $x$ over their entire range, with limiting behaviors~\cite{Schaffler:1997:p1515}
\begin{equation}
\begin{split}
    \Delta E_{\Delta_2}^\mathrm{Si} (x) &\approx -0.502 (1-x) \; \mathrm{(eV)} , \\
    \Delta E_{\Delta_2}^\mathrm{Ge} (x) &\approx 0.743 - 0.625 (1-x) \; \mathrm{(eV)} ,
\end{split}
\end{equation}
when $x\rightarrow 0$.
The linearization scheme employed here agrees well with theoretical calculations~\cite{Schaffler:1997:p1515}.

As described in Methods, for our one-dimensional model, we determine the quantum well potential $U(z_l)$ at each atomic layer position $z_l$ by linearly interpolating the conduction band offset between the barriers and the bottom of the well, yielding
\begin{equation} \label{eq:Uzl}
U(z_l) = \frac{x^d_l - x_s}{x_w - x_s} \Delta E_c,
\end{equation}
where $x_l^d$ is the Si concentration at layer $l$, averaged over the lateral probability distribution of the quantum dot wavefunction, as explained in subsection c, below. To simulate many different quantum wells, we allow for random fluctuations of $x_l^d$, due to the finite size of a quantum dot. Below, we derive the statistical distribution of $x_l^d$, as reported in Eq.~(\ref{distrbution_xld}).

\subsection{Comparison with NEMO-3D}

The two-band tight-binding model has the advantage of being computationally inexpensive, allowing us to perform many random samplings, to obtain accurate statistics. To validate the model, we compare our results to those of a more sophisticated 20-band sp\textsuperscript{3}d\textsuperscript{5}s* NEMO-3D tight-binding model (including spin)~\cite{NEMO}. NEMO-3D heterostructures are generated atom by atom. Each atom is randomly chosen to be either Si or Ge, with the probability of choosing Si given by the concentration profile $\bar x_l$ for a given atomic layer $l$. The valley splitting is computed as the energy difference between the two lowest conduction states. \\

\begin{figure*}[!ht]
	\includegraphics[width=160mm]{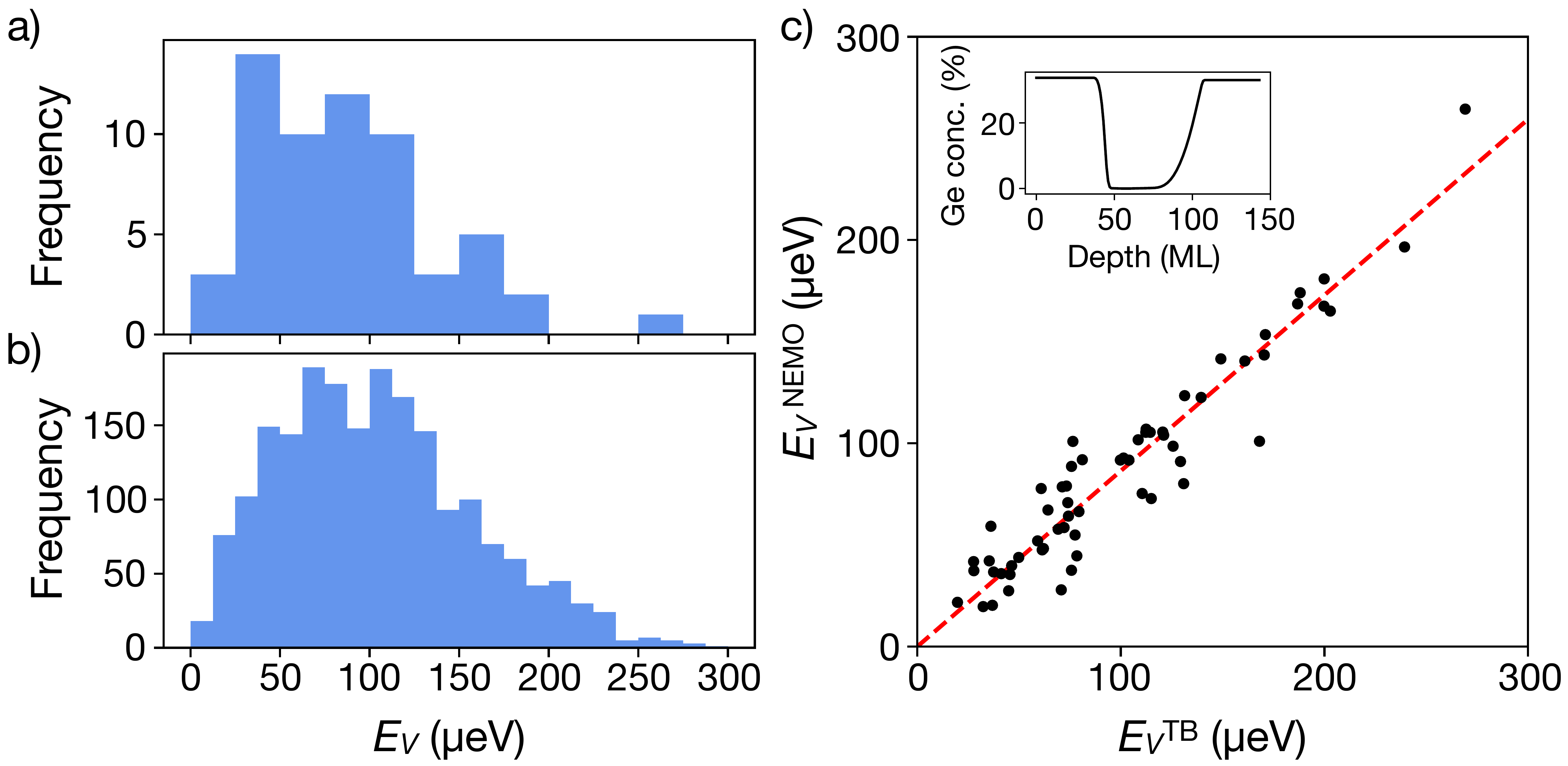}%
	\caption{A comparison of simulations with NEMO-3D and the 1-dimensional 2-band tight-binding model. (a) Histogram of 60 randomized valley-splitting simulations using NEMO-3D. (b) Histogram of 2,000 randomized valley-splitting simulations using the one-dimensional two-band tight-binding model. (c) We directly compare simulations using NEMO-3D and the 2-band model for the 60 heterostructures shown in (a). For each heterostructure, we plot its valley splitting with the 2-band model ($E_v^\mathrm{TB}$) against it's valley splitting with NEMO ($E_v^\mathrm{NEMO}$). The red dashed line shows the fit $E_v^\mathrm{NEMO} = k E_v^\mathrm{TB}$ with $k = 0.86$. All simulations assume an orbital splitting of $\hbar \omega = 2.83$~meV, and a vertical electric field of $E = 0.0125$~V/nm.}
\label{NEMO}
\end{figure*}

First, we consider the quantum well confinement profile shown in the inset of Fig.~\ref{NEMO}(c) for a quantum dot with a lateral parabolic confinement potential, corresponding to an orbital splitting of $\hbar \omega = 2.83$~meV, and a vertical electric field of $E = 0.0125$~V/nm. Note that in the two-band model, the resulting three-dimensional wavefunction is simply used to obtain a set of weighted one-dimensional layer concentrations, $x_l^d$. The results of these NEMO-3D simulations are plotted as histograms in Fig.~\ref{NEMO}(a), while the corresponding results from the two-band model are plotted in  Fig.~\ref{NEMO}(b). Although fewer random samples are obtained in the NEMO-3D case, due to computational constraints, the two distributions appear to agree well. The NEMO-3D distribution is found to have a mean of 87.4~$\mu$eV and a standard deviation of 50.1~$\mu$eV, while the two-band distribution is found to have a mean of 104.7~$\mu$eV and a standard deviation of 55.7~$\mu$eV. Qualitatively, the two distributions have similar shapes. NEMO valley splittings are on average slightly smaller, which is consistent with previous results \cite{Boykin:2004:p165325}. \\

To analyze the relationship between 2-band and NEMO simulations in more detail, we compute $E_v$ with the 1D 2-band model using the same 60 heterostructures we used in Fig.~\ref{NEMO}(a). To do so, we took the heterostructure, computed the weighted average Si concentration at each layer, and plugged it into the 1D 2-band model. We assumed the wavefunction was in the ground state of a 2D isotropic harmonic oscillator potential, with characteristic energy $\hbar \omega = 2.83$~meV. Resulting valley splittings from NEMO ($E_v^\mathrm{NEMO}$) and the 2-band model ($E_v^\mathrm{TB}$) are plotted against each other in Fig.~\ref{NEMO}(c). We can see that there is a clear, strong linear correlation between the two. Fitting these data to the relationship $E_v^\mathrm{NEMO} = k E_v^\mathrm{TB}$, we find $k = 0.86$ with standard error 0.018. Again, the fact that NEMO valley splittings are slightly less than 2-band TB valley splittings is consistent with prior results \cite{Boykin:2004:p165325}.

\begin{figure*}[!ht]
	\includegraphics[width=120mm]{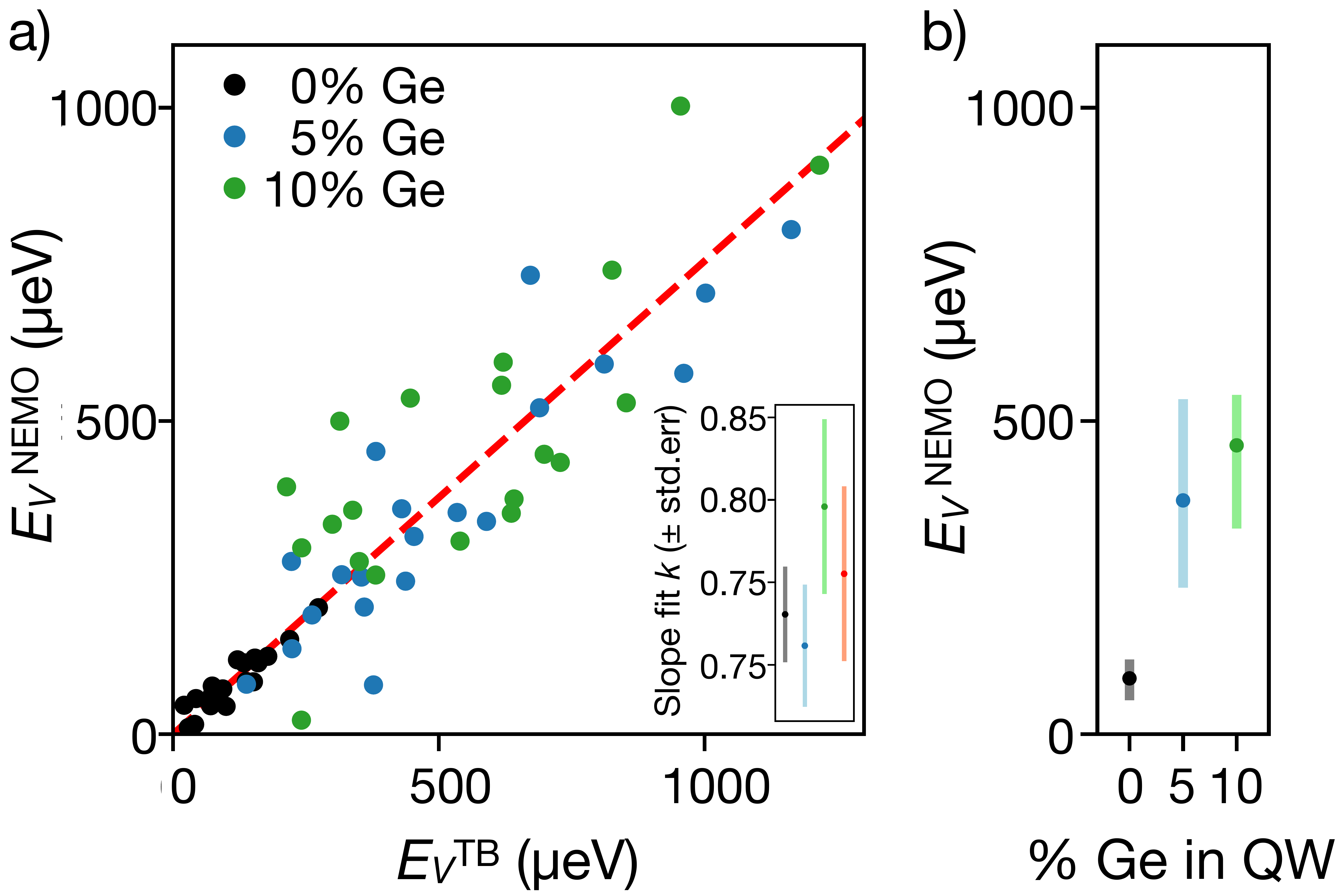}%
	\caption{(a) A comparison of valley splittings computed with the 2-band TB model ($E_v^\mathrm{TB}$) and NEMO-3D ($E_v^\mathrm{NEMO}$). Valley splittings for each point were computed using the same heterostructure on both methods. Black, blue, and green data correspond to quantum wells with 0\%, 5\%, and 10\% Ge, respectively. The red dashed line indicates the best fit $E_v^\mathrm{NEMO} = k E_v^\mathrm{TB}$, where $k = 0.76$ with standard error 0.026, shown in red in the inset. Also shown in the inset are fit values $k$ and their standard errors for each subset of data individually. (b) The mean and 25-75 percentile range of $E_v^\mathrm{NEMO}$ for quantum wells with 0\%, 5\%, and 10\% Ge. For all simulations, we use quantum wells with widths of 80 ML and interface widths of $4\tau = 10$ ML. We assume a parabolic confinement potential with strength $\hbar \omega = 2$~meV and vertical field $E = 0.0125$ V/nm. }
\label{NEMO_uniform}
\end{figure*}

Next, we analyze the relationship between 2-band and NEMO simulations as a function of the Ge content in the well. Using the same heterostructures for both 2-band and NEMO simulations, we compute valley splittings with both models, performing 20 simulations for quantum wells with each of 0\%, 5\%, and 10\% Ge. The resulting data are shown in Fig.~\ref{NEMO_uniform}(a). Again, there is a strong linear correlation between $E_v^\mathrm{TB}$ and $E_v^\mathrm{NEMO}$. This correlation is tightest for 0\% Ge, but still clear with Ge in the well. Again, we fit all the data to $E_v^\mathrm{NEMO} = k E_v^\mathrm{TB}$, finding $k = 0.76$ with standard error 0.053, indicated in the inset of Fig.~\ref{NEMO_uniform}(a). We also perform the same fit for each of the 0\%, 5\%, and 10\% Ge data individually, also shown in the inset of Fig.~\ref{NEMO_uniform}(a), and we find that the resulting $k$ values are not significantly different. We also note that these fit parameters are different than those from Fig.~\ref{NEMO}(c), indicating that there may be some interface-dependence for $k$. For these simulations, we used dots with characteristic orbital energy $\hbar \omega = 2$~meV and vertical field $E = 0.0125$ V/nm. We use 80 ML wide wells and sigmoid interfaces with widths $4\tau$ = 10 ML.\\

Fig.~\ref{NEMO_uniform}(b) shows the mean and 25-75 percentile range of the valley splittings computed with NEMO as a function of the quantum well Ge concentration. As presented in the main text, the average valley splitting clearly grows with increasing Ge content. This is a nice validation of the main results obtained with the simple 1D model.

\subsection{Statistical distribution of intervalley couplings}

Here, we examine the derivation of the statistical distribution of the intervalley coupling matrix element in more detail. From Methods, we have the intervalley coupling matrix element
\begin{equation} \label{eq:matrixElement}
    \Delta = \frac{a_0}{4} \sum_l e^{-2ik_0 z_l} \frac{x^d_l - x_s}{x_w - x_s} \Delta E_c |\psi_\text{env}(z_l)|^2 ,
\end{equation}
where the valley splitting $E_v = 2|\Delta|$. As above, $x_l^d$ represents the Si concentration in atomic layer $l$, weighted by the probability distribution of the electron charge density in the dot. We can split Eq.~(\ref{eq:matrixElement}) into its deterministic and random contributions, using the definition $x_l^d = \bar x_l + \delta_l$, where $\bar x_l$ is the ideal, smooth concentration profile of the heterostructure and $\delta_l$ are the random fluctuations about this profile. The random contribution to the matrix element can then be expressed as
\begin{equation} \label{eq:delta2}
   \delta \Delta = \frac{a_0\Delta E_c}{4(x_w - x_s)}  \sum_l e^{-2ik_0 z_l} \delta_l |\psi_\text{env}(z_l)|^2 .
\end{equation}

We now calculate the variance of the matrix element. First, we note that the random fluctuations only occur in $\delta\Delta$, so\ that $\mathrm{Var} \left[ \Delta \right] = \mathrm{Var} \left[ \delta \Delta \right]$. Second, we recall that the variance of a complex random variable is the sum of the variances of its real and imaginary components. Third, since we assume the fluctuations in different layers are independent, the variance of the sum in Eq.~(\ref{eq:delta2}) must be equal to the sum of the variances in each layer. In this way, we obtain
\begin{equation} \label{var_delta_delta}
    \mathrm{Var} \left[ \Delta \right] = \left[\frac{a_0\Delta E_c}{4(x_w - x_s)}\right]^2 \sum_l |\psi_\text{env}(z_l)|^4 \mathrm{Var} \left[\delta_l\right].
\end{equation}

To compute $\text{Var}[\delta_l]$, we note again that $x_l^d$ represents the fluctuating Si concentration in layer $l$, weighted by the electron charge density. We first describe the weighting function, which is defined in a single two-dimensional layer, and is proportional to the squared amplitude of the wavefunction. 
In the lowest-subband approximation~\cite{Friesen:2010:p115324}, the dot wavefunction is separable, so the in-plane component of the wavefunction $\psi_t(x,y)$ does not depend on the layer index. For a circular, parabolically confined dot with orbital excitation energy $\hbar\omega_\text{dot}$, the normalized wavefunction is given by $\psi_t(x,y) = (\pi a_\text{dot}^2)^{-1/2}\exp(-r^2/2a_\text{dot}^2)$, where the dot is taken to be centered at the origin, $a_\text{dot}=\sqrt{\hbar/m_t\omega_\text{dot}}$ is a characteristic dot dimension, $m_t=0.19\, m_e$ is the transverse effective mass, and $r=\sqrt{x^2+y^2}$. Let us consider the dot weighting function $w(a)$, which is defined only at the atom locations $a\in A_l$, where $A_l$ is the set of all atom positions in layer $l$. Since $w(a)$ is defined discretely, while $\psi_t(x,y)$ is continuous, their normalizations are different. To determine the normalization of $w(a)$, we require that $\sum_{a\in A_l}w(a)=1$, which is analogous to the wavefunction normalization $\int_{-\infty}^\infty dx\,dy |\psi_t (x,y)|^2 =1$. The correspondence between between the sum, $\sum_{a\in A_l}$, and the integral, $\int_{-\infty}^\infty dx\,dy$, must take into account the fact that there are two atoms in every two-dimensional unit cell of size $a_0\times a_0$, for a diamond cubic crystal like Si. Normalizing over a single unit cell, the correct correspondence is therefore given by 
\begin{equation}
\sum_{a\in A_l}\rightarrow \frac{2}{a_0^2}\int_{-\infty}^\infty dx\,dy . \label{eq:correspondence}
\end{equation}
Next, we define $w(a)=c|\psi_t(r_a)|^2$, where $c$ is a proportionality constant, to be determined below, and $\mathbf{r}_a=(x_a,y_a)$ is the 2D coordinate location of atom $a$. Using the correspondence in Eq.~(\ref{eq:correspondence}), we obtain the appropriate normalization for $w(a)$:
\begin{equation}
w(a)=\frac{a_0^2}{2\pi a_\text{dot}^2}e^{-r_a^2/a_\text{dot}^2} . \label{eq:wac}
\end{equation}

Now, according to its definition, the weighted Si concentration in layer $l$ is given by
\begin{equation}
    x_l^d = \sum_{a \in A_l} \mathbbm{1}\left[a = \mathrm{Si} \right] w(a) ,
\end{equation}
where the indicator function, $\mathbbm{1}\left[a = \mathrm{Si} \right]$, takes the value 1 if $a$ is a Si atom and 0 otherwise.
The fluctuating part of the concentration is given by $\delta_l=x_l^d-\bar x_l$. Calculating $\mathrm{Var} \left[ \delta_l \right]$, we then obtain
\begin{equation}
\label{eq:var}
\begin{split}
    \mathrm{Var} \left[ \delta_l \right] & = \mathrm{Var} \left[ \sum_{a \in A_l} \mathbbm{1}\left[a = \mathrm{Si} \right] w(a) \right] \\
    & = \sum_{a \in A_l} w^2(a) \mathrm{Var} \left[\mathbbm{1}\left[a = \mathrm{Si} \right] \right]  \\
    & = \bar x_l (1 - \bar x_l)  \sum_{a \in A_l} w^2(a)  .
\end{split}
\end{equation}
Here, in the second line, we use the fact that the atoms in the random alloy are chosen independently. In the third line, we use the fact that each indicator function in the sum is a Bernoulli trial with variance given by $p(1-p)$ \cite{bernoulliDistribution}, where the probability of success is given by $p=\bar x_l$. Making use of Eqs.~(\ref{eq:correspondence}) and (\ref{eq:wac}), we arrive at
\begin{equation} \label{eq:vardl}
\text{Var}[\delta_l]= \bar x_l (1 - \bar x_l) \frac{a_0^2}{4\pi a_\text{dot}^2} ,
\end{equation}
and
\begin{equation} \label{VarDelta}
    \mathrm{Var} \left[ \Delta \right] =  \frac{1}{\pi}\left[\frac{a_0^2\Delta E_c}{8a_\text{dot}(x_w - x_s)}\right]^2 \sum_l |\psi_\text{env}(z_l)|^4 \bar x_l (1 - \bar x_l).
\end{equation}

To complete the calculation of $\text{Var}[\Delta]$, we need to evaluate $\psi_\text{env}(z_l)$.
In this work, we compute $\psi_\text{env}(z_l)$ numerically, for the ideal concentration profile $\bar x_l$, and the corresponding confinement potential $U(z_l)$ obtained from Eq.~(\ref{eq:Uzl}), by discretizing the Schrodinger equation on the atomic lattice sites $z_l$.

Although analytical methods can be used to characterize $x_l^d = \bar x_l + \delta_l$, as in Eq.~(\ref{eq:vardl}), we also perform other types of randomized numerical simulations, as described in the main text. To do this, we could assign atoms in a 3D array, with each atom in layer $l$ having the probability $\bar x_l$ of being silicon. The statistical properties of such an array derive from a binomial distribution. In principle, for real wave functions, such calculations could involve a very large number of atoms, to accurately describe the wavefunction tails. Alternatively, we may consider a much smaller number atoms $N_\text{eff}$, for which
\begin{equation} \label{distrbution_xld}
x_l^d \sim {\frac{1}{N_\text{eff}}} \mathrm{Binom}\left(N_\text{eff}, \bar x_l\right)  ,
\end{equation}
where $\mathrm{Binom}\left(n, p \right)$ is the binomial distribution with $n$ trials and probability of success $p$.
The question now becomes, how should we determine $N_\text{eff}$? The answer is that $N_\text{eff}$ should be chosen to yield the correct statistical properties for $x_l^d$, including its mean and variance. Using the known variance of the binomial distribution, $N_\text{eff}\,\bar x_l(1-\bar x_l)$, and comparing to Eq.~(\ref{eq:vardl}), we see that we should choose $N_\text{eff}=4\pi a_\text{dot}^2/a_0^2$. This corresponds to an effective dot radius of $r_\text{eff}=\sqrt{2}\,a_\text{dot}=\sqrt{2\hbar/m_t\omega_\text{dot}}$.

\subsection{Statistical distribution of valley splittings}

Since the intervalley coupling $\Delta$ is a complex random variable and is the sum of many independent random variables (corresponding to the different layers), it follows a 2D Gaussian distribution in the complex plane.
In Eq.~(\ref{VarDelta}), we derived the variance of the intervalley coupling $\Delta$ due to concentration fluctuations. We now examine the distribution of valley splittings $E_v = 2|\Delta|$. The magnitudes of points sampled from a circular Gaussian distribution in the complex plane follow a Rice distribution \cite{Aja-Fernandez:2016}, whose probability density function is given by
\begin{equation}
    f_\mathrm{Rice} (z | \nu, \sigma) = {\frac{z}{\sigma^2}} \exp \left( - {\frac{z^2 + \nu^2}{ 2 \sigma^2}} \right) I_0 \left( \frac{z \nu}{\sigma^2} \right)
\end{equation}
where $\nu$ is the distance from the origin to the center of the circular Gaussian, $\sigma$ is the width of the Gaussian along one of its axes, and $I_0(y)$ is a modified Bessel function of the first kind. Since the Gaussian distribution for $\Delta$ is centered on the deterministic value $\Delta_0$, the probability density function for valley splittings $f_\mathrm{Rice}(E_z | \nu, \sigma)$ is centered at $\nu = 2|\Delta_0|$. The variances of the real and imaginary components are both given by $(1/2)\mathrm{Var} \left[ \delta \Delta \right]$, such that $\sigma = \sqrt{2}\sqrt{\mathrm{Var} \left[ \delta \Delta \right] }$.

\subsection{Effects of interface width and QW Ge concentration on average valley splitting}

\begin{figure*}[!ht]
	\includegraphics[width=100mm]{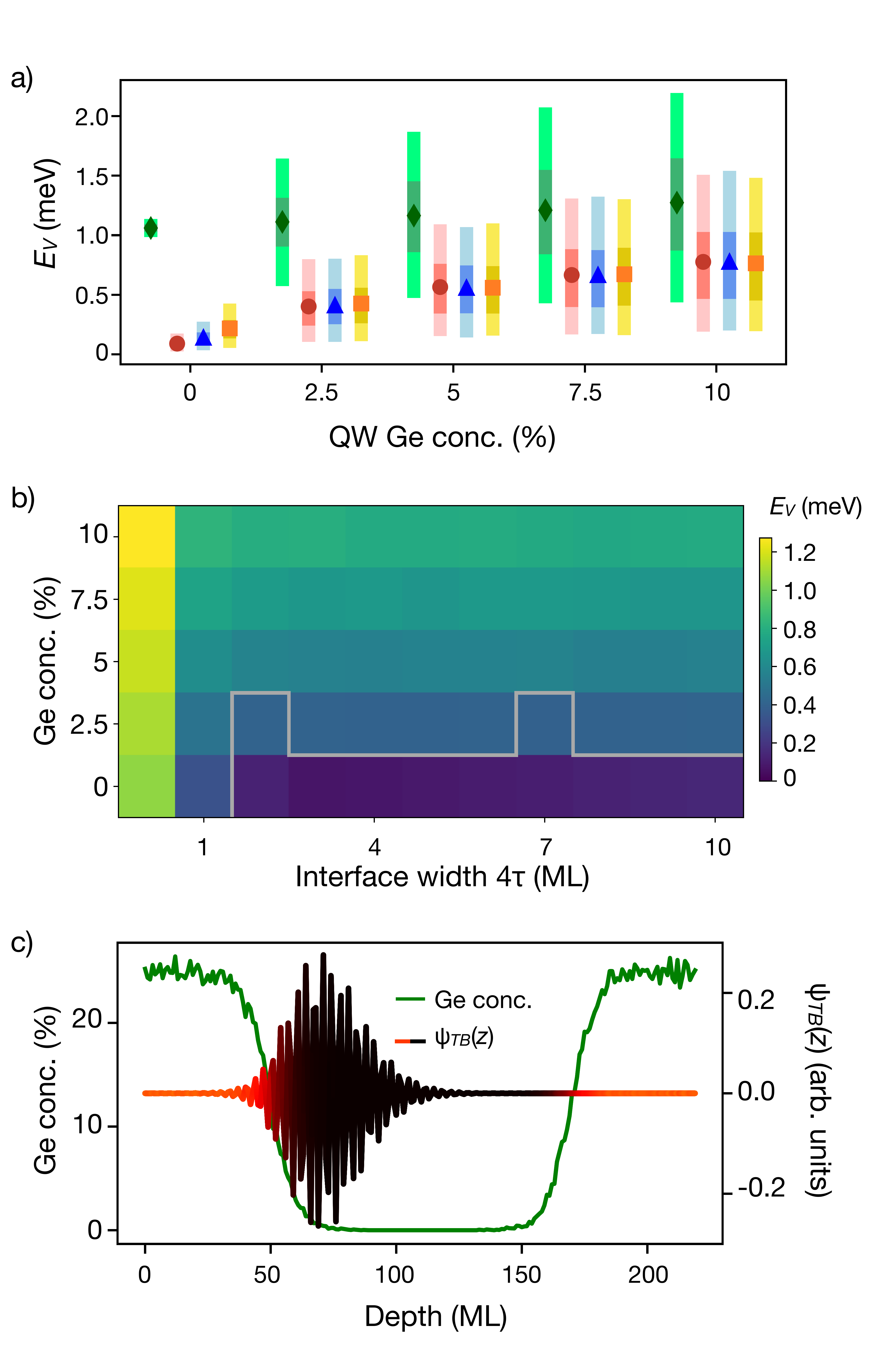}%
	\caption{Valley splitting simulations. (a) An extended version of Fig.~3(g) in the main text. Valley splitting distributions are shown for quantum well Ge concentrations from 0 to 10\% and interface widths of $4\tau = 0$ ML (green diamonds), $4\tau = 5$ ML (red circles), $4\tau = 10$ ML (blue triangles), and $4\tau = 20$ ML (orange squares). The symbol represents the mean valley splitting, while the dark bars represent the 25-75 percentile range, and the light color bars represent the 5-95 percentile range. Each bar represents 2,000 simulations using the one-dimensional two-band tight-binding model. (b) The mean valley splitting is shown for a range of interface widths $4\tau$ and quantum well Ge concentrations. Each pixel corresponds to 2,000 simulations using the one-dimensional two-band tight-binding model. All pixels above the grey line have $E_v \geq 100$~$\mu$eV in more than 95\% of the simulations; all pixels below have $E_v \geq 100$ $\mu$eV less than 95\% of the time. (c) A sample simulation of a quantum well with 0\% Ge and an interface width of $4\tau = 20$ ML. The green curve shows the Ge concentration profile, and the red-black curve shows the ground state tight-binding wavefunction. The color of the wavefunction illustrates the Ge concentration of each layer, with red corresponding to high concentrations. Clearly, a significant portion of the wavefunction is found in a region with nonzero Ge content. All simulations in this figure were performed with an electric field of $0.0075$ V/nm, for the quantum well depicted in Fig.~3(g) of the main text.}
\label{ColorPlot}
\end{figure*}

Both the average Ge concentration and the width of the interface have an effect on the valley splitting in a quantum well. Here, we analyze the contributions of both. Figure~\ref{ColorPlot}(a) provides an extended version of Fig.~3(g) in the main text, showing the valley splitting distributions at different quantum well Ge concentrations and interface widths. The valley splittings at interface width $4\tau = 0$ are consistently large, regardless of quantum well Ge concentration, due to the large deterministic component for this (somewhat unphysical) geometry. For more realistic, nonzero interface widths, increasing the Ge concentration in the quantum well increases the valley splitting. At low Ge concentration, wider interfaces can actually increase average valley splitting, because the wavefunction sees layers with more Ge. 
Figure~\ref{ColorPlot}(b) provides an extended dataset, showing the average valley splitting at several combinations of interface width vs.\ Ge concentration in the well. For very narrow interfaces, $E_v$ is large, regardless of the Ge concentration in the well. For wider interfaces, adding Ge to the well consistently boosts the valley splitting. The grey line in Fig.~\ref{ColorPlot}(b) delineates quantum wells for which $\geq$ 95\% of simulations have $E_v \geq 100$ $\mu$eV (the large, upper-left portion of the plot). For realistic electric fields, we find that any well with $>5$\% Ge, regardless of interface width, should have $E_v \geq 100$ $\mu$eV at least 95\% of the time.

Figure~\ref{ColorPlot}(c) shows a sample wavefunction in a quantum well with an interface width of $4\tau = 20$~ML. The wavefunction is colored according to the Ge concentration in each layer, illustrating how a wide interface can expose the wavefunction to more layers with nonzero Ge content. 

\subsection{Effect of vertical electric field on average valley splitting}

\begin{figure*}[!ht]
	\includegraphics[width=100mm]{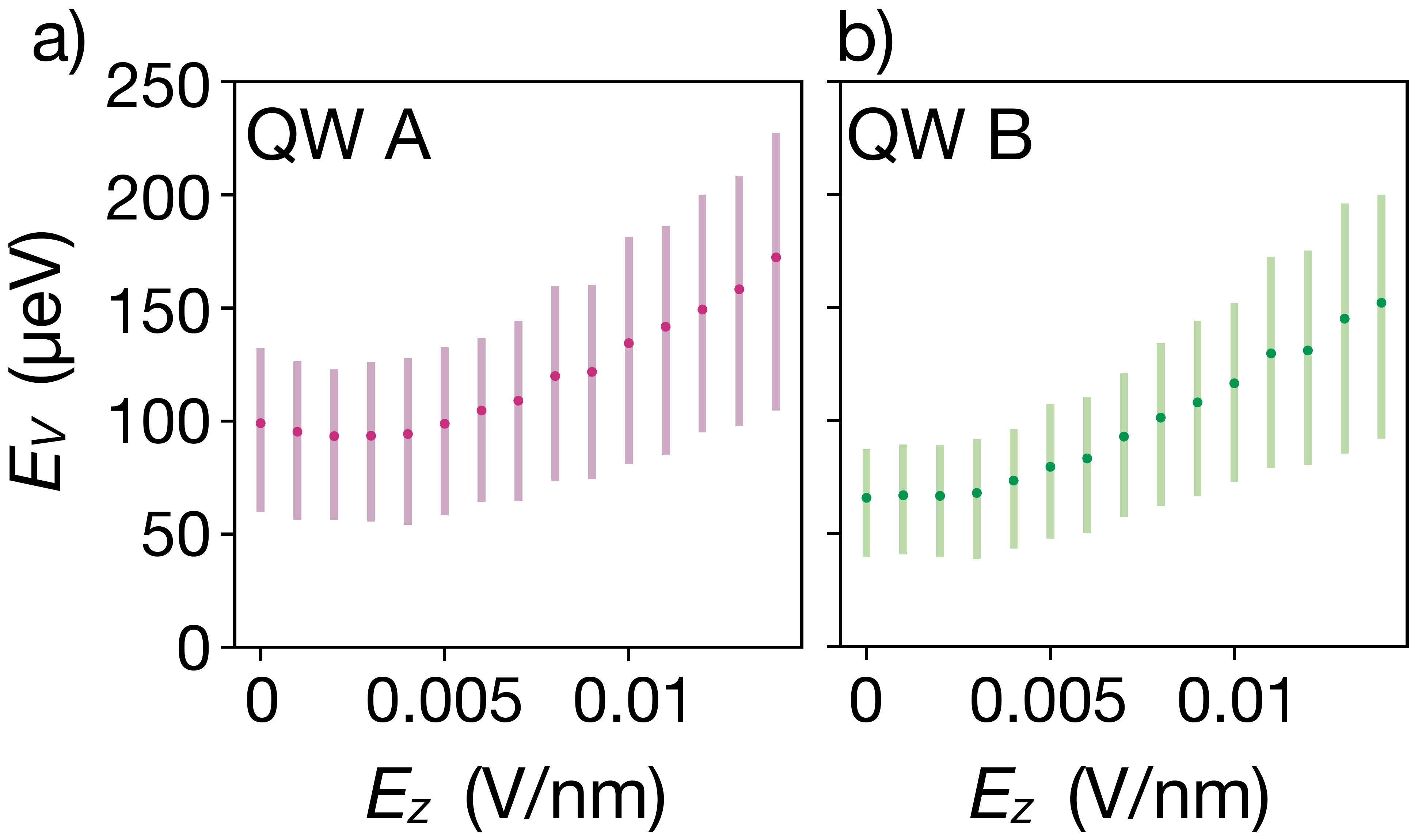}%
	\caption{The variation of the valley splitting distributions in quantum wells A (a) and B (b) as a function of the vertical electric field $E_z$. Each bar shows the mean and 25-75 percentile range of 1000 simulations using the 1D 2-band tight-binding model. We assume an orbital energy $\hbar \omega = 4.18$ meV.}
\label{EfieldVariation}
\end{figure*}

In this section, we investigate the effect of the vertical electric field on the valley splitting distributions in quantum well A and B. Fig.~\ref{EfieldVariation} shows the mean and 25-75 percentile range of 1000 1D 2-band tight-binding simulations of $E_v$ with various vertical fields $E_z$, for both quantum wells A and B. Increasing the vertical field leads to larger mean $E_v$ and larger spreads in $E_v$ as the quantum dot wavefunction penetrates the top interface, thereby increasing the alloy disorder. That said, even wells with zero vertical field still show a sizeable spread in $E_v$ due to alloy disorder. In this paper, we use $E_z = 0.0075$ V/nm because the resulting $E_v$ distributions agree well with the data.

\clearpage
\newpage
\bibliography{supplement.bib}